\documentclass[screen,sigplan, nonacm]{acmart}

\copyrightyear{2024}
\acmYear{2024}
\setcopyright{cc}
\setcctype{by-nc}
\acmConference[ASPLOS '24]{29th ACM International Conference on Architectural Support for Programming Languages and Operating Systems, Volume 4}{April 27-May 1, 2024}{La Jolla, CA, USA}
\acmBooktitle{29th ACM International Conference on Architectural Support for Programming Languages and Operating Systems, Volume 4 (ASPLOS '24), April 27-May 1, 2024, La Jolla, CA, USA}
\acmDOI{10.1145/3622781.3674180}
\acmISBN{979-8-4007-0391-1/24/04}

\usepackage[]{hyperref}

\everypar{\looseness=-1}

\begin{document}

\title{Toleo: Scaling Freshness to Tera-scale Memory using CXL and PIM}

\author{Juechu Dong}
\email{joydong@umich.com}
\orcid{0000-0002-8855-9962}
\affiliation{%
  \institution{University of Michigan}
  \city{Ann Arbor}
  \state{Michigan}
  \country{USA}
}

\author{Jonah Rosenblum}
\email{jonaher@umich.edu}
\orcid{0009-0006-4755-9575}
\affiliation{%
  \institution{University of Michigan}
  \city{Ann Arbor}
  \state{Michigan}
  \country{USA}
}

\author{Satish Narayanasamy}
\email{nsatish@umich.edu}
\orcid{0000-0001-5016-1214}
\affiliation{%
  \institution{University of Michigan}
  \city{Ann Arbor}
  \state{Michigan}
  \country{USA}
}

\begin{abstract}

Trusted hardware's freshness guarantee ensures that an adversary cannot replay an old value in response to a memory read request. They rely on maintaining a version number for each cache block and ensuring their integrity using a Merkle tree. However, these existing solutions protect only a small amount of main memory (few MBs), as the extraneous memory accesses to the Merkle tree increase prohibitively with the protected memory size. We present Toleo, which uses trusted smart memory connected through a secure CXL IDE network to safely store version numbers. Toleo eliminates the need for an unscalable Merkle tree to protect the integrity of version numbers by instead using smart memory as the root of trust. Additionally, Toleo ensures version confidentiality which enables stealth versions that reduce the version storage overhead in half.

Furthermore, in the absence of Merkle tree imposed constraints, we effectively exploit version locality at page granularity to compress version number by a factor of 240. These space optimizations make it feasible for one 168 GB Toleo smart memory device to provide freshness to a 28 TB CXL-expanded main memory pool in a rack server for a negligible performance overhead. We analyze the benefits of Toleo using several privacy-sensitive genomics, graph, generative AI, and database workloads.

\end{abstract}

\maketitle 

\section{Introduction}

\label{sec:intro}
Confidential computing~\cite{ccc2022} protects the privacy and integrity of code and data on untrusted third-party systems such as the public cloud. It enables mutually untrusting users to pool and process sensitive data without compromising their privacy, which is vital for advancing applications such as trustworthy AI~\cite{MLTEE, FederatedLearningTEE}, population-scale health analytics~\cite{AES-XTS}, and collaborative databases~\cite{enclaveDB2018priebe}. It is also useful in edge devices for protecting intellectual property (e.g., ML model parameters) and ensuring computation integrity even when an adversary has physical access to those devices. 

Confidential computing uses trusted hardware~\cite{ccc2022} as the root of trust to provide a trusted execution environment (TEE). Intel SGX~\cite{costan2016sgxexplained}, for instance, supports secure enclaves, which isolate code and data during an application’s execution from the rest of the system, including the operating system, and administrators with physical access to the system. More specifically, it provides three guarantees: \textbf{confidentiality} ensures privacy of data and code, \textbf{integrity} ensures that code and data are not tampered with, and \textbf{freshness} ensures that a memory read returns the most recent value which defends against replay attacks~\cite{IntelMEE}.

{\bf Guaranteeing freshness is expensive}, and no prior solution has shown that it is feasible to protect memory sizes at tera-scale.  The original SGX provided freshness only for 128 MB of secure main memory called the enclave page cache (EPC). A modern data-intensive application with a working set larger than 128 MB experiences frequent page faults to swap data in and out of the EPC. Studies have reported 5x slowdown for certain applications~\cite{taassori2018vault}, which will only increase with memory size. With the advent of tera-scale memory systems enabled by Compute Express Link (CXL) technology, which are critical to support applications like large language models (LLMs), there is an acute need for large secure memory.  It is unlikely that the current Merkle tree-based approach to guarantee freshness could scale from MBs to TBs, which are six orders of magnitude larger. 

Given these challenges, processor manufacturers dropped freshness protection from their latest products in exchange for improved memory scalability. Intel’s recent scalable SGX~\cite{johnson2022supporting}, for instance, does not guarantee freshness, which is a significant security compromise and hinders our progress towards zero-trust systems.

{\bf The key bottleneck in providing freshness is the use and maintenance of Merkle tree}~\cite{merkle1987digital}. Each cache block is associated with a non-repeating version number. This version helps check if a memory read’s return value is the most recent version or not. However, version numbers are also stored in memory and and are prone to replay attacks and tampering. The Merkle tree is introduced to solve this problem. Each tree node holds meta-data to guarantee the integrity and freshness of its children. The leaves hold the data cache block’s version numbers. By ensuring that the tree’s root is always stored within the trusted CPU, freshness is guaranteed for all the versions, and therefore all the data. 

To check the freshness of a memory read, the freshness of all nodes from the root to the leaves must be checked. A version number cache helps mitigate this overhead, but research has shown that even with 32 KB of cache per core, the hit rate is low (60-70\%)~\cite{umar2022softvn}. For 128 MB cache and 8-ary tree, this can incur up to 7 additional accesses per read/write, but it increases to 13 accesses for 28 TB memory. Also, the version cache hit rate suffers as the tree size grows. Recent work proposed solutions to reduce the size of the integrity tree~\cite{taassori2018vault,saileshwar2018morphable, umar2022softvn} and support up to 64 GB of secure memory. However, the fundamental scalability challenges associated with the use of Merkle trees remain.

\label{revis:Intro-InvisiMem}
Smart memory (2.5D~\cite{25D,25Dcost} or 3D~\cite{HBM, HMC}) integrates logic close to DRAM, and tampering with connections will destroy the package. Prior work such as InvisiMem\cite{aga2017invisimem} and Obfusmem\cite{awad2017obfusmem} expanded the trusted computing base (TCB) to include logic in smart memory to mitigate address side-channel. They also argue that storing all data in use on trusted smart memory eliminates the need for freshness checks and, consequently, the need for a Merkle tree to maintain versions. However, these solutions store all in-memory data on smart memory devices, which is expensive for tera-scale memory systems.  

In contrast, Toleo enables freshness checks at scale. Toleo uses a small trusted smart memory (168 GB) to store only versions, which replaces the role of Merkle trees for maintaining versions. This allows us to scale freshness checks by nearly six orders of magnitude to a tera-scale memory pool (28 TB) in a rack server. Toleo achieves this goal by leveraging two emerging technologies: {\bf smart memory and CXL}.  By including this logic within the trusted computing base (TCB), Toleo creates a trusted memory and uses it to store version numbers. By construction, trusted Toleo memory guarantees confidentiality, integrity, and freshness for version numbers when they are stored off-chip. CXL 2.0 Integrity and Data Encryption (IDE) link helps guarantee these properties during off-chip transmission to Toleo. 

Toleo stores a version number per cache block, while the data and MAC tags are stored in conventional DRAM. SGX’s version numbers (56 bits) incur significant space overhead (1:9.1 meta-data to data ratio). Smart memory is expensive, so we propose two solutions to significantly reduce version space requirements.

One is {\bf Trip (Tri-level Page) format}. It takes advantage of version locality -- adjacent addresses tend to have the same versions or they only differ slightly. Most data-intensive workloads exhibit significant spatial locality for writes, translating to version locality. Previous work~\cite{umar2022softvn,leetnpu,nacommoncounters,hua2022mgx,taassori2018vault,saileshwar2018morphable} exploited this program behavior to compress the Merkle tree. While the Merkle tree structure imposes restrictions on this optimization, our simple one-level version list enables new opportunities for efficient representation. Our Trip format efficiently represents versions of all cache blocks within a page using three different formats that are dynamically chosen based on the version locality of a page. In the common case, we need as little as 12B to protect a 4 KB page, resulting in excellent space efficiency (1:341), which is about 38 times more efficient than using a full version number. 

We also exploit Trip’s page-level format to simplify and improve the efficiency of the version cache. We propose to {\bf extend the last-level TLB’s entry}~\cite{bhattacharjee2011tlb} with 12B to store the corresponding page’s version numbers in the Trip format. This solution improves {\bf version cache hit rate to 98\%} on average, which is significantly higher than 60-70\% hit rates reported for caching Merkle Tree.

Our second solution is short {\bf stealth versions}, which are half the size of the original versions. Version numbers serve two purposes: to protect freshness and to serve as nonces for the AES block cipher. The nonce must be unique to guarantee strong confidentiality of the AES cipher. 

We observe that the non-repeating property of version numbers is essential for nonces, but not strictly needed to provide a strong enough freshness guarantee. Therefore, we propose to split a large version (64-bits in our design) into two parts: upper-version (UV) and stealth version. UV consists of 37 high-order bits, and stealth consists of the remaining lower-order bits. 

We show that it is sufficient to store only short stealth versions within Toelo to guarantee freshness with an adequate cryptographic margin. Short stealth versions can repeat during a program’s execution. Therefore, we observe that ensuring the confidentiality of stealth versions is essential, unlike in the Merkle Tree design. While Toleo and CXL IDE, by construction, guarantee confidentiality, we need to ensure that stealth versions cannot be inferred from public information such as addresses. We solve this problem using a simple but efficient solution to randomize the stealth versions. A key aspect of our design is that it achieves this randomization without compromising version locality.

Through these space optimizations, we reduce version space overhead significantly (1:240 meta-data to data).  As a result, we show that just one shared 168 GB Toleo smart memory connected through CXL 2.0 is sufficient to protect an entire 28 TB of main memory (aggregated capacity of CXL-expanded local memory and shared memory pool) in a 4-node rack-server where each node has 128-cores. The performance overhead of guaranteeing freshness on top of confidentiality and integrity is negligible (2\%). Even though Toleo is located on a remote drawer and shared across all the nodes in a rack server, due to our excellent version cache hit rate (98\%), the latency and bandwidth impact of fetching stealth versions is marginal.

\section{Motivation and Background}

\subsection{Threat Model} 
\label{sec:threat}

We assume the standard threat model of trusted hardware like Intel SGX, which assumes a powerful adversary with control of the middleware (OS and hypervisor). The adversary also has physical access to the machine and can probe and tamper with the communication in the off-chip network, including the DDR and CXL channels. The adversary, however, cannot observe or modify any communication of the on-chip interconnect, TSV (Through-Silicon Via) in 3D memory, interposers in 2.5D memory, or any microarchitectural states enclosed in a silicon package.

We assume a trusted processor with support for attestation and isolated execution of enclaves such as Intel SGX. We also assume a smart memory with trusted compute logic. We assume both of these devices' microarchitectures are correct and protected against side-channels~\cite{oleksenko2018varys, brasser2019dr}. We assume Toleo can easily track write frequencies and perform rate limiting if it detects a Rowhammer threat~\cite{yauglikcci2021blockhammer}. Lastly, we assume that the application running inside the secure enclave is written correctly, in that it guarantees all outputs do not leak privacy.

If an integrity or freshness check fails, it triggers a "kill-switch", where the processor logs an error, destroys the enclaves along with the sensitive data, and shuts down.

\subsection{Trusted Hardware for TEE}
\label{sec:bg:tee}

Trusted execution environments (TEE) protect data, code, and the execution of confidential processes or virtual machines on untrusted computing systems. TEEs rely on trusted hardware to provide necessary security guarantees and enable confidential computing applications~\cite{MLTEE, GenomeTEE, FederatedLearningTEE}. 

Toleo focuses on improving SGX trusted hardware as it has the strongest threat model. The original SGX implementation, or "client SGX", seeks to provide three guarantees for code and data stored in main memory. Confidentiality ensures privacy. Integrity ensures data and code cannot be tampered with (modified). Freshness guarantees load-store semantics for memory, wherein a memory read to an address returns the last value written to it. To protect data privacy, all secure memory used by enclave applications is encrypted by the SGX memory encryption engine (MEE). 

Confidentiality is guaranteed by decrypting the data only within the trusted processor. On a last-level cache eviction, data is encrypted before it is written back to memory. Client SGX uses AES counter-mode encryption. It uses a secret key $k$, a non-repeating (nonce) version $v$, and a plain-text $p$ as input to create a cipher-text $c$, or $AES_{CTR}(k, v, p) = c$. The nonce ensures that two writes to an address with the same value always produce different cipher texts.

Intel SGX Scalable uses AES-XTS~\cite{AES-XTS} instead which does not require a nonce, and only uses a secret key to encrypt plain text $AES_{XTS}(k, p) = c$. As a result, two writes to an address with the same data will yield the same cipher text. While this does not allow an adversary to break the encryption, it does partially violate confidentiality as an adversary can perform traffic analysis to infer same-value writes.

Integrity is guaranteed using message authentication code:

\begin{equation*}
\begin{gathered}
MAC = Hash_{key}(Version, address, cipher)
\end{gathered}
\end{equation*}

On each memory read, this MAC is recalculated to verify that the response has not been modified. Without the secret key, an adversary cannot generate valid MACs.

Freshness ensures memory read-write semantics: memory read to an address returns the value of the most recent write to that address. Freshness prevents replay attacks, where an adversary snoops a response to a prior memory read to an address and replays that old response. This defeats the MAC integrity checks, and could potentially modify control flow, violate memory safety, and induce leaks. Thus, replay attacks can violate the confidentiality and integrity of data.

To guarantee freshness in hardware, Client SGX uses monotonically increasing nonces, or version numbers, per cache block. The MAC integrity check, whose hash depends on version numbers, would succeed only if the response is for the latest version. The problem with this approach is that the freshness of the version number itself is not guaranteed, which is why a Merkle tree is needed.

As noted in the introduction, a Merkle Tree solution is expensive, especially to support tera-scale memories. To overcome this issue, Intel released "scalable SGX" using AES-XTS. Scalable SGX avoids MAC and Merkle tree-related overheads by abandoning freshness and integrity protection.

Table~\ref{tab:sgxcompare} summarizes the differences in guarantees provided by client and scalable SGX. There is need for a client SGX-like solution that provides all three guarantees, including freshness for the entire tera-scale main memory for a low overhead. This paper seeks to address this need. 

\begin{scriptsize}
\begin{table}[h]

\begin{tabular}{|l|l|l|l|}
\hline
Protects   & 
\begin{tabular}[c]{@{}l@{}} Client  SGX\end{tabular} & 
\begin{tabular}[c]{@{}l@{}} Scalable SGX\end{tabular} & 
\begin{tabular}[c]{@{}l@{}}Toleo\end{tabular} 
\\ \hline \hline
Full Physical Memory Space & No      & Yes        & Yes        \\ \hline
Confidentiality    & Yes       & Partial    & Yes        \\ \hline
Integrity          & Yes                 & No       & Yes         \\ \hline
 Freshness          & Yes           & No         & Yes       \\ \hline
\end{tabular}
\caption{Memory Protection Comparison}
\label{tab:sgxcompare}

\end{table}
\end{scriptsize}

\section{CXL IDE and Smart Memory}
\label{sec:tech}

\subsection{Compute Express Link (CXL)}
\label{sec:cxl}

The Compute Express Link (CXL) \cite{cxlspec} is an open-source standard providing high-bandwidth, low-latency connections on PCIe links between devices like accelerators and memory expanders. Unlike PCIe drivers, CXL ports provide memory semantics, reducing latency and granularity issues.

A key use of CXL is expanding server memory via CXL memory expanders\cite{CXLMemorypool}.  CXL 2.0 enables main memory expansion from local DRAM to larger, disaggregated pools\cite{pond}. 

Figure~\ref{fig:rack} shows a typical rack-level memory pooling solution with 4 compute nodes. Each 128-core compute node is attached to its local DRAM (3 TB). It is also connected to a shared large CXL-attached memory pool (16 TB) in another drawer. Trusted hardware-enabled TEE needs to protect data and code stored in this entire large (28 TB) main memory.

CXL 2.0 introduces Integrity Data Encryption (IDE) \cite{CXLIDESKID}, providing secure, end-to-end connections between CXL ports on trusted components. It ensures confidentiality, integrity, and replay protection at the flit level using non-deterministic AES stream ciphers and MAC checks. IDE SKID mode enables near-zero latency and low bandwidth overhead by processing data without waiting for integrity checks. In particular, we operate on the selective IDE stream to accommodate any switches on the link.

The CXL IDE supports the TEE Device Interface Security Protocol (TDISP)\cite{tdisp}, which provides two main functions for our design: 1) Establishing a trust relationship between the smart memory and the host processor. 2) Enabling key exchange and establishing a secure channel on CXL. Additionally, it offers the flexibility to securely attach/detach the smart memory to/from a Trusted Virtual Machine.

\subsection{Smart Memory}
\label{sec:bg:smart}

3D and 2.5D packaging technologies connect multiple dies via high throughput interconnects and pack them into one heterogeneous package. These technologies provides high-density, smart memory solutions such as 3D-stacked High Bandwidth Memory (HBM) \cite{HBM} and Hybrid Memory Cube (HMC) \cite{HMC}, which stacks 16 layers of 32Gb DRAM \cite{Murdock2022} atop a logic layer, connecting via Through-silicon-vias (TSV). A single package can accommodate up to four 3D-stacks\cite{Aquabolt-XL}. Conversely, 2.5D packaging \cite{25D, 25Dcost} offers a more cost-effective method, connecting multiple chiplets on the same plane using silicon interposers with TSV. The in-package TSVs make it nearly impossible for adversaries to tamper with or observe communication without damaging the device. 

Process In Memory (PIM) systems, leveraging such technologies, can offer up to 168GB of memory space \cite{UPMEM}. Smart memories are increasingly popular due to their enhanced memory bandwidth, essential for big-data workloads like LLMs. However, in our work, we utilize smart memory not primarily for bandwidth advantages but to develop a trusted memory solution. 

\label{revis:SmartMem}
Prior works \cite{valamehr3Dcontrolplane, gu3dsecurity, aga2017invisimem, awad2017obfusmem} have utilized smart memory to enhance hardware security. Specifically, InvisiMem\cite{aga2017invisimem} expanded the trusted computing base (TCB) to smart memory devices to mitigate memory address and timing side-channels. 
Invisimem replaces all passive DRAM with smart memory and utilizes the integrated logic in memory to build a secure communication channel between the smart memory and the trusted processor. This secure channel conceals memory addresses from adversaries and eliminates the need for freshness checks. However, considering the high cost of 2.5D/3D packaging technologies, replacing all main memory devices in a rack (28 TB of capacity) with smart memory would be prohibitively expensive. In contrast, Toleo enables freshness checks, instead of eliminating them. Toleo uses smart memory only to store versions, whereas data still lives in conventional memory. Combined with our Trip version compression format and partial stealth versions, Toleo achieves a smart memory to data size ratio of 1:240. This allows us to share a single small smart memory  device of 168GB across multiple nodes and protect all 28TB of conventional memory in a rack.

\section{Design}

This section introduces Toleo, a system that utilizes a compact 168 GB smart memory and CXL technology to securely manage memory freshness across a rack-scale system with 28 TB of physical memory. 

\begin{figure}[h!]
    \includegraphics[width=\linewidth]{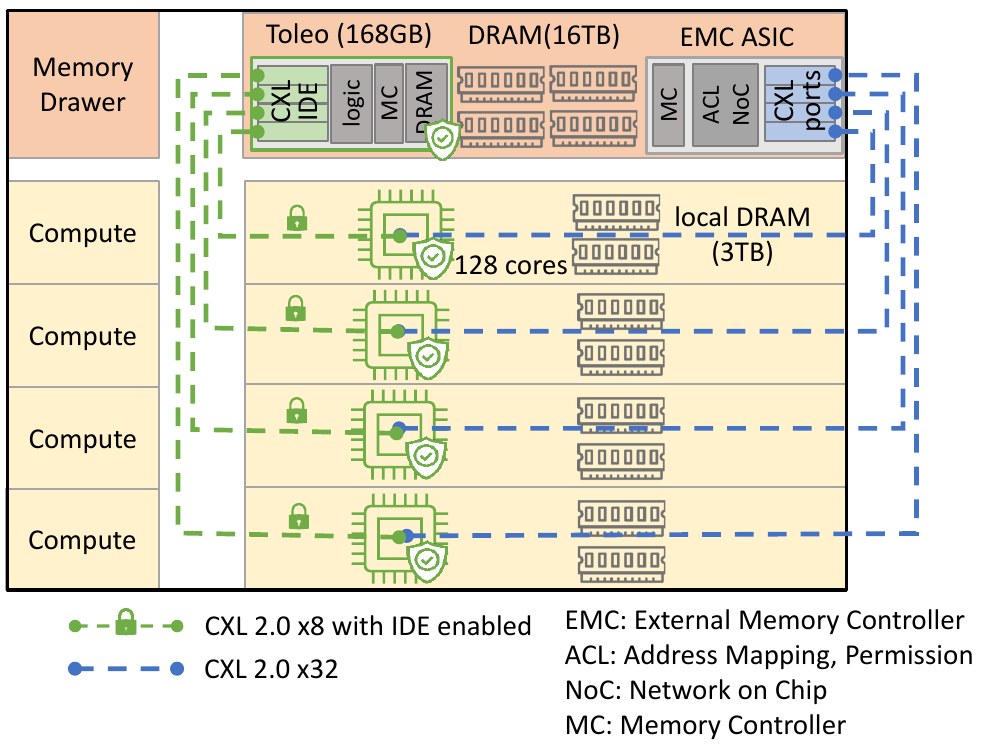}
    \caption{168 GB Toleo protects 28 TB memory in a rack.}
    \label{fig:rack}
\end{figure}

\subsection{Toleo  - Leveraging Smart Memory and CXL for Versions}
\label{sec:smartmemory}

\begin{figure}[h!]
    \includegraphics[width=0.8\linewidth]{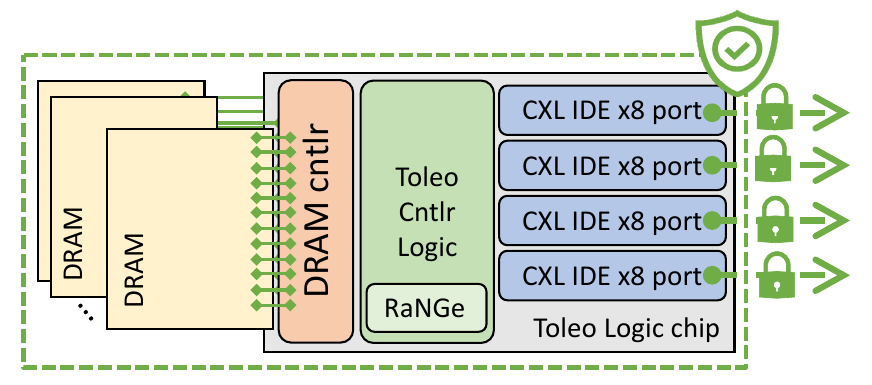}
    \caption{Toleo Design}
    \label{fig:toleo}
\end{figure}

Toleo (Fig. \ref{fig:toleo}) is a trusted smart memory device. The logic layer near the memory is trusted. It consists of CXL IDE port, DRAM controller, version management logic that includes a simple general-purpose in-order core, a DRAM-based random number generator (RaNGe\cite{RaNGe}), and a private hardware embedded attestation key. As covered in Section~\ref{sec:bg:smart}, the connections between this trusted logic layer and the DRAM modules are package-enclosed and impenetrable. We propose that Toleo be used as the root of trust for securely storing stealth version numbers and provide memory freshness guarantees for the entire rack.

{\bf Trusted memory obviates the Merkle Tree.} Like client SGX, we store data, and MAC tags in conventional memory. Each cache block is assigned a version number, which is used to provide confidentiality and freshness. However, unlike SGX, version numbers are stored in trusted Toleo, which obviates the need for a Merkle tree and its overheads to verify the integrity of version numbers retrieved from memory (Section~\ref{sec:intro}). During boot time, the host's trusted CPU authenticates Toleo, and thereafter uses it to store stealth versions for cache blocks it accesses. As a trusted piece of hardware, Toleo's, its responses adhere to load-store semantics, ensuring the return of the most recent value (stealth version) for the requested address.

{\bf CXL IDE enables secure stealth version transmission.} The trusted CPU connects to Toleo via CXL, as shown in Figure~\ref{fig:rack}. CXL IDE provides an attestation protocol for Toleo authentication and ensures the confidentiality, integrity, and freshness of stealth versions during host-Toleo transfer. It uses a non-deterministic stream cipher, so identical transactions yield different cipher texts. As discussed later, this feature enables us to use short stealth versions. Additionally, CXL's skid mode~\cite{CXLIDESKID} introduces negligible latency and bandwidth overhead for security checks by allowing the release of version numbers before integrity check completion and enabling parallel computation of memory security primitives and CXL IDE checks. We avoid speculative attacks due to this parallelization by withholding data from the CPU until both checks are done.

{\bf Sharing a Toleo device across the rack through CXL.} CXL expanders allow multiple host CPU nodes to share a large 16 TB pool of DRAM, as detailed in Section~\ref{sec:cxl}. Combined with each host's local memory (3 TB per node), this totals 28 TBs of memory in need of protection. This capacity is six orders of magnitude greater than what conventional client SGX can feasibly protect.

Considering smart memory is more expensive~\cite{PIMHotChip33} than conventional memory, a compact version number representation is crucial for cost-effectiveness. Compact version numbers also improves the efficiency of on-chip caching.

We propose optimizations that make it feasible for a single 168 GB Toleo smart memory device to protect the entire 28 TB memory pool in a rack. As depicted in Figure~\ref{fig:rack}, one Toleo connects to four compute nodes. This connection divides its 32-bit CXL links into four separate 8-bit links, with one link dedicated to each host.

In this section we will present three solutions to address the space, latency, and bandwidth challenges of sharing Toleo across a rack. First is a short "stealth version", a small (27-bit) version counter, which is about half the size of conventional version numbers (56-bits). The second solution is compressing Stealth versions at the page granularity using our Trip (Tri-level Page) format. Finally, we present a stealth version cache that achieves a 98\% hit rate and helps reduce the impact on memory latency and bandwidth due to sharing Toleo.  

\subsection{Partial Stealth Versions}
\label{sec:design:stealth}

In client SGX, the version number serves two purposes. The first purpose is as a nonce for the AES block cipher. The block ciphRer requires a non-repeating nonce for every encryption. It is vital that the version number does not repeat, as repetitions enable an attacker to infer the secret AES key.~\cite{IntelMEE}. 
SGX uses 56-bit versions, which is sufficient to ensure non-repeatability during a system's lifetime. In Toleo, we use 64-bit versions and ensure the non-repeatability property (discussed below).

The second purpose of the version number is to protect freshness, as explained in Section~\ref{sec:bg:tee}. We observe that Toleo does not need to store the entire version number in smart memory to guarantee freshness.  Only the {\bf partial stealth version} (27-bits), which is the lower part of the 64-bit full version, needs to be stored in Toleo. The upper version (UV) consisting of the 37 most significant bits is stored in conventional memory. This observation further reduces Toleo's space overhead.

Our insight is that by protecting the confidentiality, integrity, and freshness of stealth versions, we can provide a strong freshness guarantee for data with fewer bits. We ensure that the stealth versions stay confidential from end to end, and thus a replay attack can't do any better than randomly replaying an old version of a cacheline and hoping that the stealth version happens to match the latest. Given that the stealth version comprises 27 bits, the probability of a successful match is $1$ in $2^{27}$, and an adversary is limited to a single attempt due to the 'kill-switch' discussed in Section~\ref{sec:bg:tee}.

Now that we have established that confidential stealth version numbers enable freshness, we will explain how their confidentiality is guaranteed. 

Toleo's smart memory guarantees confidentiality for stealth versions when they are stored off-chip. CXL IDE’s non-deterministic stream-cipher encryption ensures confidentiality when stealth versions are communicated between Toleo and the trusted CPU. 

In addition, it is crucial to prevent an adversary from inferring a stealth version using observable three public information: ciphertext,  MAC, and address. AES-XTS block cipher ensures that cipher-text and MAC do not leak the tweak (64-bit nonce + address). 

{\bf Address side-channel}: If the stealth version starts at a known value (e.g., 0) and increases sequentially as with client SGX, an adversary can easily infer it from the number of writes to an address. 

We address the above problem by initializing the stealth version to a random value from the range $0$ to $2^{27}$. It increments monotonically and wraps over at overflows until our reset policy (discussed below) decides to reset the stealth version to a different random intial value. The random initial value for the stealth version breaks its address trace dependency, which helps preserve confidentiality.

At a stealth version reset, its corresponding upper version  (UV) of the full version is incremented.  Thus, although the partial stealth versions may repeat, the full versions will still be unique, so long as the stealth versions are reset before they increment past their randomly initialized start value. 

A naive solution is to reset the stealth version number when it is one less than the initial random value. This guarantees a reset and UV increment before a stealth version is reused. However, this requires that we maintain the initial stealth value to determine when to reset, in addition to the current value. Thus,  the naive solution doubles the number of bits needed,  defeating the purpose of only storing a partial stealth version in Toleo. 

Instead, on each stealth version increment, we reset it with a small probability ($1$ in $2^{20}$). With 27-bit stealth versions, the odds that a reset does not happen before we loop back to its initial random value are $1.7 \times 10^{-19}$ (see Sec. \ref{sec:security:nonce} for details), which is exceedingly unlikely.

{\bf Summary:} In summary, our version number has 64 bits in total. We protect stealth versions' (27-bit) confidentiality, integrity, and freshness using Toleo smart memory attached via CXL IDE. The upper versions (37-bit UV) are stored in conventional memory. To prevent adversaries from inferring stealth versions using public address trace, we randomize an initial value for a stealth version when it is reset. Finally, we randomly reset with a small probability on every increment to reduce space, while also ensuring the non-repeatability of the full version with an exceedingly high probability. We analyze the security implications of these design choices further in Section~\ref{sec:security}.

\subsection{Trip: Page-level Stealth Version Compression}
\label{sec:compression}

To further reduce Toleo's space overhead, we seek to take advantage of version locality: adjacent addresses tend to have the same versions or they only differ slightly. This is largely due to a high degree of spatial locality exhibited by writes, especially in modern data-intensive workloads. 

\begin{figure}[h]
    \includegraphics[width=\linewidth]{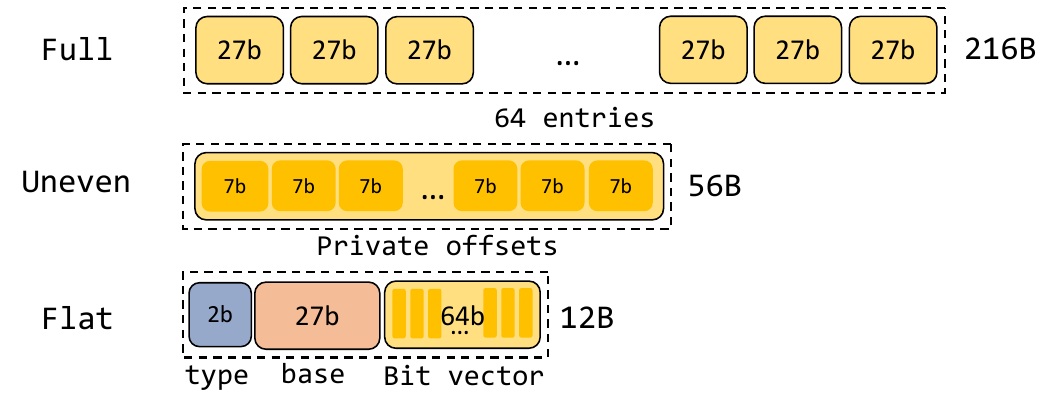}
    \caption{Stealth Version Entries}
    \label{fig:version-entries}
\end{figure}

{\bf Version locality:} Modern big-data workloads exhibit a high degree of version locality. They often write to large data structures uniformly, even though they may be read arbitrarily. Consider the example of an intermediate layer of a large language model. The intermediate layer is initialized and then updated multiple times for each token generation. Each update makes the corresponding cache lines dirty in the on-chip cache system. The version numbers are updated when they get evicted from LLC, thus the version numbers for each cache line in a page differ at most by a constant stride of one. Similarly, dynamic programming arrays for genome alignment are updated uniformly. 

In addition, some data structures are read-only or write-once. For example, edge lists for graph kernels and multi-level hash tables for genomic analysis kernels are read-only once initialized. This means neighboring cache lines in the physical memory space often share the same version number and are updated close in time. 

{\bf Solution}: Unlike past solutions to compress version numbers~\cite{taassori2018vault,saileshwar2018morphable}, we propose Trip (tri-level page) format to compress version numbers at the page granularity. We assume a 4 KB page containing 64 cache lines, each of size 64 bytes.  Stealth versions of cache blocks within a page are stored in one of three formats depending on the degree of version locality they exhibit: flat, uneven, and full (Fig. \ref{fig:version-entries}). Trip format is expected to have good compression ratios unless the workload frequently evicts dirty cache lines from LLC to arbitrary locations in memory. 

{\bf Flat: }  Flat entries are used for pages with a high degree of version locality, including read-only pages, write-once pages, and pages with a uniform write where adjacent addresses' versions differ by one. In the data-intensive workloads that we study, these pages constitute 92\% of all pages.

Each physical page in the memory system is statically mapped to a flat entry. Initially, all pages use flat versions. A flat entry contains one shared stealth version and a 64-bit vector, where each bit corresponds to a cache block within that page. For each page, the stealth version is initialized with a random value as explained in Section~\ref{sec:design:stealth}, and its bit-vector is set to 0. When a cache block is updated, its corresponding entry in the bit-vector is set. When the entire bit-vector for a page is set, its shared stealth version number is incremented and the bit-vector is reset. 

A flat entry provides an excellent data-to-metadata ratio (1:341) compared to a naive representation that stores the full 27-bit stealth for each cache block (1:19).

{\bf Uneven:} When the stride between stealth versions for cache blocks within a page exceeds one, we allocate an uneven entry in addition to a flat entry. The uneven entry contains a 7-bit offset for each of the 64 cache blocks, which accommodate strides up to 128. The last 48 bits of the flat entry's bit-vector field store a pointer to the uneven entry in Toleo's physical memory space. Thus,  the flat entry continues to be used in this format as well.

Toleo keeps track of the maximum (MAX) and minimum (MIN) of uneven offsets in the upper 14 bits of the flat entry bit vector. When an offset overflows its 7-bit value, Toleo normalizes private offsets by subtracting the MIN from all the offsets of the uneven entry and adding it to the base.

Uneven entries have a compression ratio of 1:60, and thus help manage pages that are written unevenly yet still exhibit significant version locality.
 
{\bf Full:} Finally, if the stride exceeds 128 (\( 1/2^{7}\)), then Toleo deallocates the uneven entry and allocates a full entry. The corresponding flat entry now points to the full entry. The full entry is an uncompressed list of 27-bit stealth versions, each corresponding to a cache block within the page. 

{\bf Stealth Reset:} As we described in Section~\ref{sec:design:stealth}, the stealth version of a cache block could be reset on an increment with a $\frac{1}{2^{20}}$ probability. Since Trip manages versions at page granularity, we check to reset when the ``leading'' cache block with the highest version number within a page is incremented. For flat, the first cache block that gets written after the last increment to the shared stealth base has the leading version. For uneven, MAX offset tracked in flat's bit-vector is the leading version. For full, we use the 27b base in the flat entry to keep track of the leading version. Any increment to these leading versions causes Toleo to check for a stealth reset. 

{\bf UV compression:} We maintain one shared UV for all cache blocks within a page. MACs organized at cache block granularity have spare space, which we leverage to store this information. When a stealth version is reset, the corresponding shared UV is incremented, and all the cache blocks within the page are re-encrypted with the new version number in addition to generating new MACs. While re-encrypting the whole page is an expensive operation it is a rare operation as resets happen with a low probability ($\frac{1}{2^{20}}$), and the cost of re-encryption is amortized across millions of writes.

{\bf Page free and remap:} Toleo allows the operating system (OS) to downgrade a page's entry to flat when it is freed due to remapping or deallocation. During this downgrade process, Toleo resets the stealth version and increments the UV version without re-encrypting the data, which effectively scrambles the page content. Because the version number is used in calculating the Message Authentication Code (MAC), attempting to read the old page contents right after we reset the version number will cause the MAC to be calculated incorrectly and thus the integrity check will fail. This mechanism safeguards against any malicious intent of the OS to downgrade an active page to read its confidential contents. Moreover, while a malicious OS may fail to downgrade pages after freeing them, this would only result in suboptimal space utilization, posing no security risk.

In scenarios where Toleo exhausts its available space, it is the responsibility of the host OS to ask Toleo to downgrade inactive pages to flat. If Toleo is full, it will reject update requests until sufficient space has been freed through downgrade requests.

\subsection{Metadata Memory Layout and Caching}
\label{sec:cache}

 \begin{figure}[h]
    \centering
    \includegraphics[width=\linewidth]{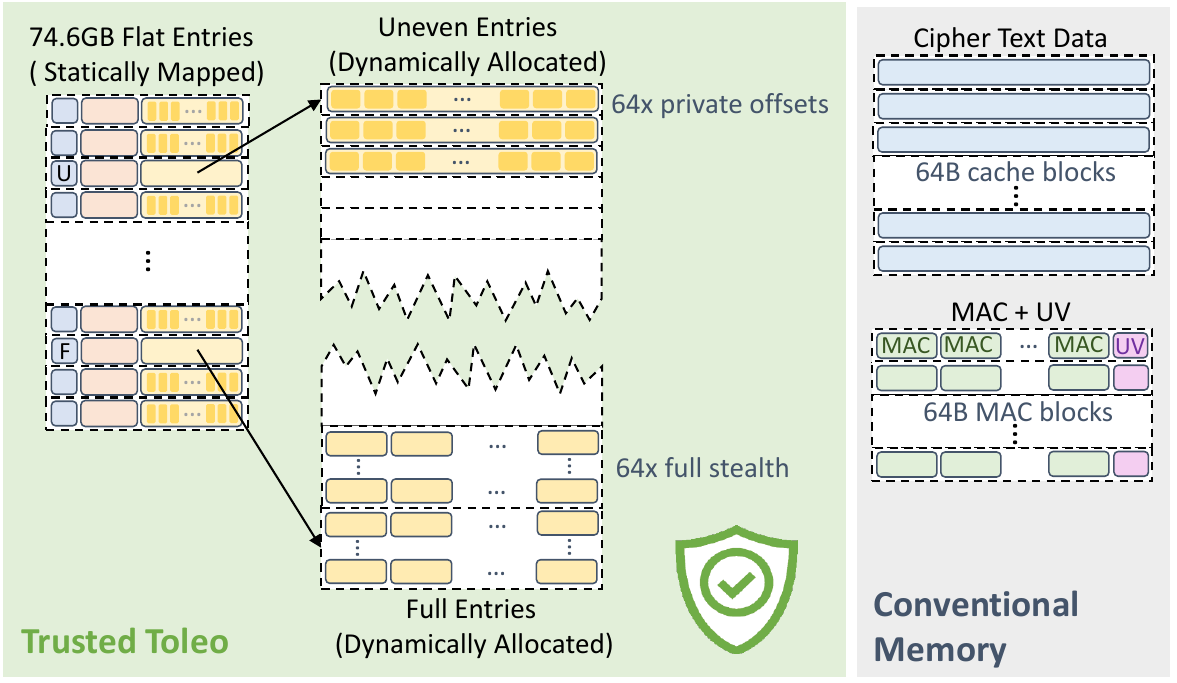}
    \caption{Memory layout for data, MAC, and versions.}
    \label{fig:meta-data-layout}
\end{figure}

 \begin{figure}[h]
    \centering
    \includegraphics[width=\linewidth]{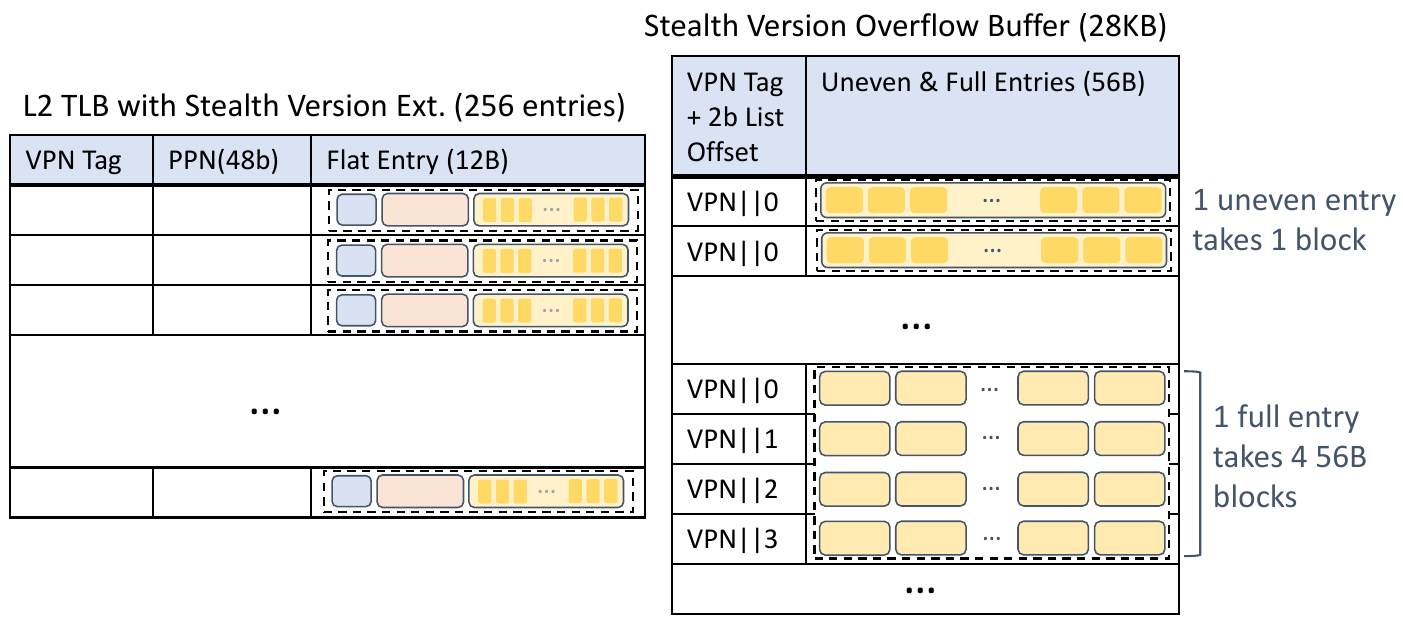}
    \caption{Caching Stealth Versions. }
    \label{fig:stealth-cache}
\end{figure}

{\bf Memory Layout:} We store cipher-text data, MAC tags, and Upper Versions (UVs)  in conventional memory. Eight MACs (each 56 bits) are packed into a cache block and stored on the same memory node as data as shown in Figure~\ref{fig:meta-data-layout}. This MAC block has spare space which we use to store the shared UV corresponding to those cache blocks. This is possible as UV is shared across all the cache blocks within a page. This design choice also eliminates the need for extra memory accesses to fetch UV. 

The physical memory space on the rack, totaling 28 terabytes (TB), is partitioned such that 24.8TB is dedicated to storing encrypted data, and the remaining 3.2TB is allocated for MACs and UVs. 

Toleo uses a static flat-entry array and two dynamic lists of uneven and full entries that grow toward each other. The flat entry array, occupying 74.6GB, contains an entry for every physical page protected by Toleo (24.8TB in our study). The remaining 93.4GB of Toleo's free space is allocated dynamically for uneven and full entries as needed. When a page is represented in the uneven/full format, the flat entry's bit-vector acts as a pointer to identify the specific location of its corresponding uneven/full entry.

{\bf Meta-data Caches:} Stealth versions are cached in trusted host CPU  as shown in Figure \ref{fig:stealth-cache}.  Our version caches are inclusive. When there is a miss in the Last Level Cache (LLC), these two caches are accessed simultaneously. 

Each flat entry corresponds to a page. Therefore, we cache them in an extended last-level Translation Lookaside Buffer (TLB)\cite{LLTLB}.  The extension does not change the tag array of the TLB. The data array is extended with an additional 12 bytes (flat-entry size). For a 256-entry, this amounts to just 3KB. 

Uneven and full stealth versions reside in a 28KB stealth version overflow buffer. This buffer's block size matches uneven-entry size (56-byte).  A full version entry thus consumes four blocks, identified with a 2-bit offset combined with the Virtual Page Number (VPN) to form the tag.

Like client SGX, the MAC blocks are cached in a dedicated 16-way, 32KB/core MAC cache located on the trusted processors. UV, which is co-located with MAC in a block,  is cached along with the MAC.

\section{Implementation}

The Toleo accepts 3 types of requests from the host processor: {\tt READ}, {\tt UPDATE} and {\tt RESET}. The first two are translated into CXL.mem READ/WRITE transactions, whereas the third is a specific CXL.mem WRITE to a Memory Mapped Register(MMR).  {\tt READ} and {\tt UPDATE} both return stealth versions, but the latter also increments the version. {\tt RESET} is used by the host OS to downgrade an entry to flat. The host trusted processor accepts {\tt UV\_UPDATE} from Toleo when it resets a stealth version. It increments the UV and re-encrypts the page with the new version number.  To downgrade a Toleo entry after freeing or remapping a page, the host OS requests a {\tt UV\_UPDATE} and a Toleo {\tt RESET}.   

The Toleo controller has a simple in-order general-purpose core that is programmed to handle the three requests.  A D-RaNGe\cite{RaNGe} random number generator serves as the source of randomness for re-initialization.

\section{Security Analysis}
\label{sec:security}

We assume SGX's threat model (Section~\ref{sec:threat}). 
Three design choices in Toleo are critical to its security. One is trusting smart memory to ensure CIF (confidentiality, integrity, and freshness) of stealth versions stored in it. The second is constructing the AES-XTS nonce as a concatenation of monotonically increasing upper version (UV) and randomized stealth version.  
The third is protecting the freshness of only the partial stealth version (27-bit) of a full version (64-bits). We will analyze the security guarantees of each choice below.

\subsection{Smart Memory for Storing Versions}
\label{sec:smart-secure}

We assume that the physical links in smart memory connecting trusted logic to the memory modules are tamper-proof, unlike off-chip links that connect a trusted processor to conventional memory. 
In a smart memory device the logic die and DRAM die are 2.5D/3D stacked and enclosed in the same silicon package. These connections are tiny $\mu$-bumps (<30 $\mu m$) or TSV (5$\mu m$) that are sandwiched between two layers of silicon. These are much smaller (50x - 300x) than conventional memory bus connections between CPU and main memory. Additionally, one cannot observe these connections without undoing the silicon packaging and attaching wires to a bare die. Doing this “surgery” inevitably destroys the chip's functionality, making smart memory tamper-proof. Leveraging the TDISP protocol supported by CXL IDE, the trusted smart memory can establish a secure channel with the host via CXL IDE and safely communicate stealth versions.

\subsection{Full Version Is Non-Repeating}
\label{sec:security:nonce}

Client SGX used a monotonically increasing 56-bit version per address to ensure it does not repeat even if one assumes a serial process of $2^{56}$ updates that takes 8 years of continuous processing on a dedicated platform \cite{IntelMEE}. We effectively use a 64-bit full version number, divided into UV and stealth versions. Therefore, our version will also not repeat provided we ensure that the stealth version never repeats in a "stealth interval", which is the interval between two UV increments.

Full version collision only happens when UV is not incremented for $2^{27}$ consecutive updates to the same address and exhausts the 27-bit stealth version space. We establish that the probability of stealth space exhaustion is less than $1.7\times10^{-19}$ in a sample space of $2^{56}$ updates.

Consider $2^{56}$ continuous updates to the same physical address. Divide these updates into $2^{30}$ continuous stealth intervals, each with $2^{26}$ updates. The chance of no reset in a given interval is $(1-2^{-20})^{2^{26}} = 1.6 \times 10^{-26}$. The chance of no reset in any of the intervals is $(1-1.6\times10^{-26})^{2^{30}} = 1.7 \times 10^{-19}$, which is extremely low. There cannot be $2^{27}$ continuous updates without reset if all $2^{30}$ stealth intervals have at least one reset, therefore full version collision is very unlikely. 

\subsection{Security of Partial Stealth Versions}

 An adversary cannot snoop or forge a stealth version while it is stored in trusted smart memory, or when it is communicated to the CPU  over the secure channel (Section~\ref{sec:smart-secure}). An adversary must identify an old transaction (Ciphertext, MAC, and UV) with the same stealth version as the current pending transaction, and to successfully replay it.  However, since partial stealth versions (27-bits) are confidential, an adversary only has a ($2^{-27}$) chance of a successful replay attack.  If a replay attempt fails, the system will immediately shut down and prevent further attempts.

Toleo and CXL IDE guarantee the stealth version's confidentiality as stated earlier. In addition, we now establish that stealth versions cannot be inferred from public memory traces (cipher text, MAC, address, and UV).

\textbf{Memory traffic analysis}: AES-XTS provides confidentiality for its tweak (version) and plain-text data unless the tweak repeats along with the data value. As established earlier, our version is statistically guaranteed to be non-repeating, and therefore it defeats memory traffic analysis attacks based on cipher texts and MACs. The address trace is not dependent on the stealth version. Also, as we randomize stealth's initial value, it is independent of address trace. Finally, while UV increments are dependent on stealth, the stealth version is reset after an increment, and therefore future stealth versions are independent of the UV state.

\textbf{Toleo Timing Side-channels}: Two of our optimizations, Trip format for compression, and version caching, do not affect the cryptographic infrastructure but can affect timing. Timing side channels can infer Toleo's actions including version read/update, entry upgrade/downgrade/reset. These actions are only dependent on the memory update traces and OS requests. They are independent of the random seed used for the stealth version initialization. For example an uneven entry upgrades to a full entry when its 7-bit private offset overflows, which is independent of its 27-bit base derived from the random reset seed. Therefore, Toleo timing side channels do not leak any information that is not already known by the adversary.

\section{Evaluation}
\label{sec:evaluation}

We evaluate Toleo using cycle-accurate simulation of 12 privacy-sensitive big-data applications from several benchmark suites: GenomicsBench~\cite{genomicsbench}, GAP Benchmark Suite~\cite{GAP},  llama2.c~\cite{touvron2023llama}, and in-memory databases redis~\cite{redis}, memcached~\cite{memcached}, and hyrise~\cite{hyrise}.  Banded Smith-Waterman ({\tt bsw}) and {\tt chain} are 2D/1D dynamic programming workloads with a large input data size. FM-Index Search ({\tt fmi}), De-BruijnGraph Construction ({\tt dbg}), and Pileup Counting ({\tt pileup}) have irregular memory access patterns from tree traversal, hash table access, and graph construction. Page Rank ({\tt pr}), Breath-First Search ({\tt bfs}), and Single-Source Shortest Path ({\tt sssp}) are graph algorithms with irregular memory footprints. {\tt llama2-gen} is a large language model inference workload dominated by matrix multiplications. 

{\tt redis} and {\tt memcached} are in-memory key-value store databases. They handle all-write key-value store requests generated by the memtier benchmark\cite{memtier} following a Gaussian distribution. {\tt hyrise} is a high-performance in-memory SQL database and runs the online transaction benchmark TPC-C\cite{tpc-c}. 

Their peak resident set size (RSS) and last level cache (LLC) miss rate per kilo instructions (MPKI) are listed in Table \ref{tab:bench}.

\begin{table}[h!]

\centering
\resizebox{\linewidth}{!}{
\begin{tabular}{|l|l|l|l|l|l|}
\hline
\multicolumn{1}{|l|}{\textbf{bench}}  & \multicolumn{1}{|l|}{\textbf{LLC mpki}} & \textbf{RSS}     & \multicolumn{1}{|l|}{\textbf{bench}}      & \multicolumn{1}{|l|}{\textbf{LLC mpki}}   & \textbf{RSS}     \\ \hline\hline
\multicolumn{3}{|c|}{\textbf{GenomicsBench}}                       & \multicolumn{3}{|c|}{\textbf{Graph - GAP Suite}}                                 \\ \hline
\multicolumn{1}{|l|}{bsw}    & \multicolumn{1}{|l|}{1.21} & 11.7GB  & \multicolumn{1}{|l|}{bfs}        & \multicolumn{1}{|l|}{22.57}  & 12.9GB  \\ \hline
\multicolumn{1}{|l|}{chain}  & \multicolumn{1}{|l|}{0.49} & 11.75GB & \multicolumn{1}{|l|}{pr}         & \multicolumn{1}{|l|}{133.98} & 20.8GB  \\ \hline
\multicolumn{1}{|l|}{dbg}    & \multicolumn{1}{|l|}{0.47} & 9.86GB  & \multicolumn{1}{|l|}{sssp}       & \multicolumn{1}{|l|}{2.41}   & 24.57GB \\ \hline
\multicolumn{1}{|l|}{fmi}    & \multicolumn{1}{|l|}{0.45} & 12.05GB & \multicolumn{3}{|c|}{\textbf{Generative AI (LLM)}}                                  \\ \hline
\multicolumn{1}{|l|}{pileup} & \multicolumn{1}{|l|}{0.66} & 10.85GB & \multicolumn{1}{|l|}{llama2-gen} & \multicolumn{1}{|l|}{57.96}  & 25.8GB  \\ \hline
 \multicolumn{6}{|c|}{\textbf{In memory Database}}\\ \hline 
 redis& 0.76\footnotemark{} & 11.8GB& memcached& 3.14&11.8GB\\ \hline
 hyrise& 3.14& 6.96GB& & &\\\hline
\end{tabular}
}
\caption{Benchmarks.}
\label{tab:bench}
\end{table}
\footnotetext{For Redis, LLC MPKI is calculated for the core running the main thread.}

We use cycle-accurate CPU simulator SniperSim \cite{sniper} integrated with DRAM simulator DRAMSim3\cite{DRAMsim} to conduct performance analysis. We instrumented benchmarks with PinPlay\cite{looppoint}  to fast forward, warm up, and then simulate a region of interest of 100 million instructions per core for performance analysis. We run long simulations of 32 billion instructions after warming up the cache with 3.2 billion instructions to evaluate the Toleo usage for dynamically allocated uneven/full entries. We calculate statically allocated flat entry usage from the peak RSS reported by the OS. 

\begin{scriptsize}
\begin{table}[h]
  \centering
  \small
\resizebox{\linewidth}{!}{
  \begin{tabular}{|l|l|}
    \hline
    \textbf{Processor}  & \textbf{2.25 GHz, 32 cores}\\
    \hline
    Cores                & 6-way wide dispatch, 320-entry RoB\\
    \hline
    L1-I/D cache        & 32 KB per core\\ \hline 
                        & 8-ways, 4 cycles latency, LRU\\
    \hline
    L2 Cache            & 1 MB per core \\ \hline 
                        & 16-ways, 14 cycles latency, LRU\\
    \hline
    L3 Cache            & 16MB shared by every 8 cores \\ \hline 
                        & 16-ways, 49 cycles latency, LRU\\
    \hline
    DRAM                & DDR4-3200  256GB/channel, 3 ch \\
    \hline\hline
    \textbf{CXL Mem. Pool}      & 16TB (shared with other nodes), 1TB available \\
    \hline
    CXL2.0 (w re-timer)   & PCIe5.0 8x (12.7GB/s BW, 95ns lat.)\cite{IntelCXL,pond}\\
    \hline
    DRAM               & DDR4-3200, 2 channels\\
    \hline
    \hline
    \textbf{Mem. Protection Engine} &  \\
    \hline
    AES                 & 40 cycle latency, 1 per cycle throughput \\
    \hline 
    MAC Cache           & 1 MB (32 KB per core), 16-way, LRU \\
    \hline
    L2 TLB Stealth Ver. Ext.         & 256 entries total, fully associative   \\
    \hline
    Stealth Ver. Overflow Buf.  & 28 KB (512 entries) total, 16-way, LRU     \\
    \hline
    \hline
    \textbf{Toleo}             &  168GB (shared with other nodes)   \\
    \hline 
    CXL2.0 IDE (w/ re-timer)   & PCIe5.0 2x (3.32GB/s BW, 95ns lat.) \cite{IntelCXL,pond}\\
    \hline
    DRAM               & HMC2, 16B transactions, 15ns lat. \\
    \hline\hline
 \textbf{InvisiMem}&\\
 \hline
 DDR Channels&DDR4-3200, 3 ch\\\hline
 Local DRAM&
6x HMC2, 32 links, 32B transactions\\ \hline 
 CXL Mem. Pool DRAM&4x HMC2, 32 links, 32B transactions\\ \hline
  \end{tabular}%
}
\caption{Simulation Configuration \protect\footnotemark}
\label{table:zen4}
\end{table}
\end{scriptsize}
\footnotetext{To construct a reasonable workload for the mostly single-threaded Redis, we run it on a 1/3 down-scaled configuration with 11 cores, 1/3 cache size, 1/3 CXL/AES bandwidth, and 1 DDR4-3200 channel to simulate 3 instances of Redis running in parallel.}

We model four configurations. One is a baseline with no memory protection (NoProtect). Second (CI) adds confidentiality (C) through AES-XTS encryption and integrity (I) by checking MAC. It is equivalent to scalable SGX's Total Memory Encryption (TME) with an integrity guarantee added. CI does not guarantee freshness (F). Third is Toleo, which adds freshness using our CXL-enabled smart memory design.
Lastly we compare Toleo against InvisiMem\cite{aga2017invisimem}-far, an all-smart memory design that mitigates memory address and timing side-channel in addition to CIF. 

\begin{figure*}[ht]
\centering
\begin{tabular}{@{}c@{}@{}c@{}}
\includegraphics[width=0.44\linewidth]{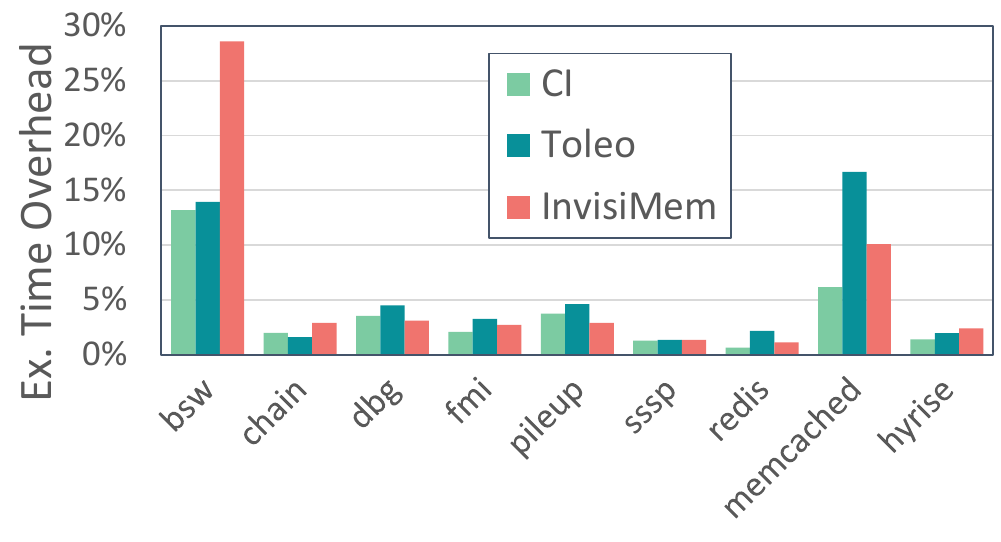} &
\includegraphics[width=0.23\linewidth]{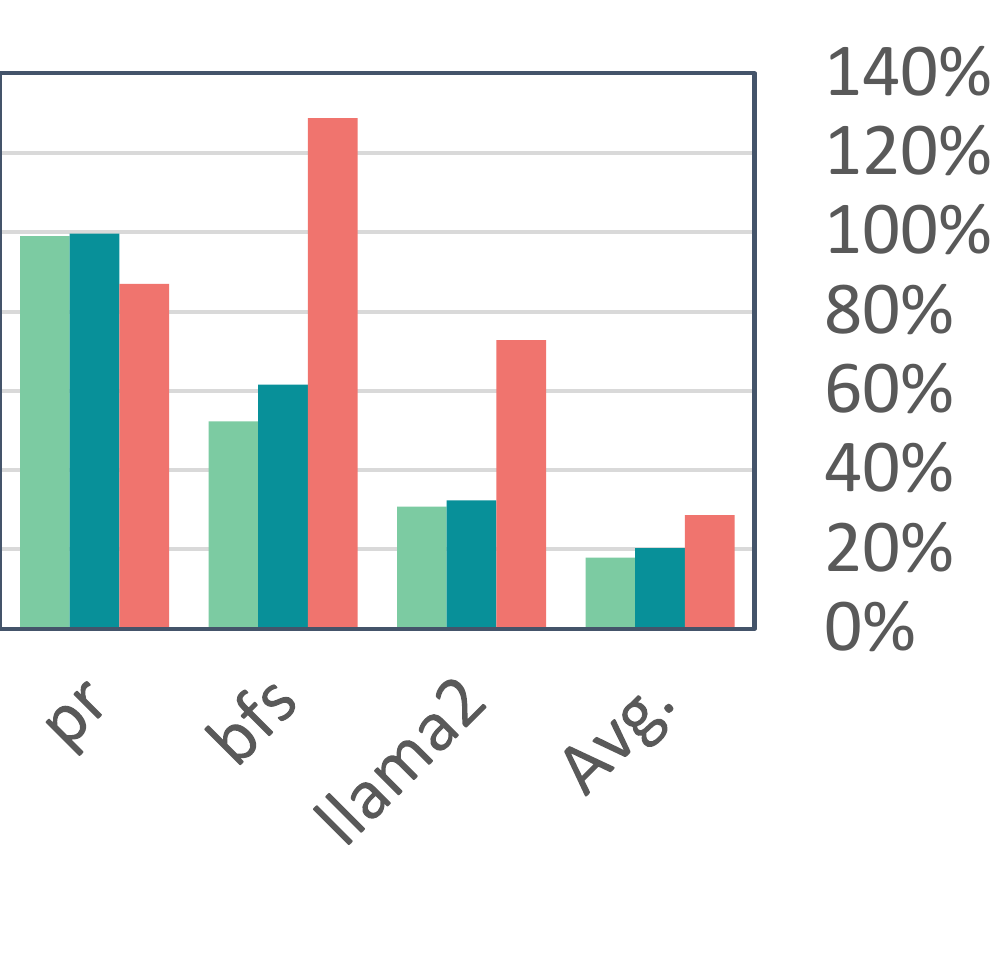} \\
\end{tabular}
\caption{CI and Toleo Performance Overhead.}
\label{fig:performance}
\end{figure*}

We model a 1/4th down-scaled compute node based on AMD EPYC 9754 processor\cite{EPYC} with 32 cores and 768GB local memory as shown in Table \ref{table:zen4}. 
The compute node is connected to a 16TB memory expansion pool (1TB available to the compute node) through 8x CXL2.0 PCIe5.0 links, and a 168GB Toleo through 2x IDE-enabled CXL2.0 PCIe5.0 connection. The CXL links are equipped with a re-timer to accommodate the need to connect to memory devices shared by the whole rack (distance \textgreater500mm)\cite{pond}. Virtual pages are randomly mapped to the remote memory pool and local DRAM proportional to the CXL and DDR bandwidth to optimize for memory bandwidth utilization.  We insert an AES module, a MAC cache, the L2 TLB stealth version extension, and a 28KB stealth version overflow buffer between the memory system and LLC. For the InvisiMem setup, local DRAM and CXL memory expander DRAM are modeled as two HMC2 modules per DDR channel.

\subsection{Performance Analysis}
\label{sec:performance}

Figure \ref{fig:performance} shows the performance overhead for CI, Toleo, and InvisiMem compared to no memory protection. CI incurs an average overhead of 18\%. For most applications, CI's overhead is relatively small (<15\%), but it is higher for memory bandwidth-intensive workloads such as {\tt pr}, {\tt bfs}, and {\tt llama2}. Toleo only adds 1-2\% overhead to provide freshness on top of CI. An exception is the memory latency-sensitive workload {\tt memcached}, which incurs an additional 11\% overhead. Toleo serves 28 TB of main memory, which is about six orders of magnitude larger than 128 MB of EPC in client SGX's Merkle tree-based solution. InvisiMem incurs a higher overhead than CI and Toleo, averaging 29\%. This is due to InvisiMem's mechanisms that provide two additional security guarantees against memory address and memory bus timing channels. They require encrypting messages twice,  forcing write and read packets to be of the same size, and sending dummy packets over the memory bus to ensure a constant rate of communication.

\begin{figure}[h!]
\centering
\includegraphics[width=\linewidth]{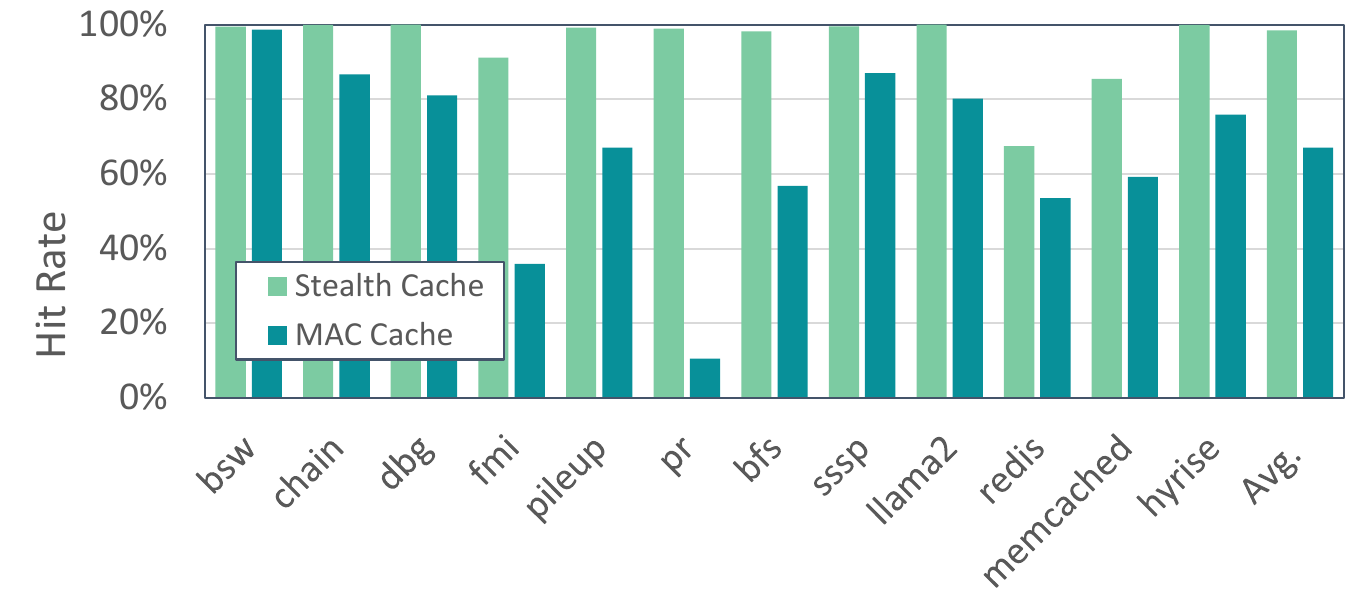}
\caption{ Cache Hit Rates}
\label{fig:cache-hit}
\end{figure}

To understand CI, Toleo, and InvisiMem's performance overhead, we analyze meta-data cache hit rates, memory bandwidth, and latency overheads. Toleo's negligible overhead is primarily due to the high hit rates to its meta-data caches. Stealth versions are maintained at the page granularity in a highly compressed format (Section~\ref{sec:compression}). As discussed later (Section~\ref{sec:eval:meta-date-size}), we reduce the stealth-to-data ratio to 1:240. This enables effective caching. Stealth's flat versions are cached along with L2 TLB, which exhibits high hit rates due to spatial locality. In the uncommon case, a stealth overflow buffer is used to cache uneven and full entries. We observed only compulsory misses to our 28 KB stealth overflow buffer in our simulation window. The two stealth caches together enjoy a high hit rate of 98\% on average as shown in Figure~\ref{fig:cache-hit}. Notably, the in-memory key-value store workloads \texttt{redis} and \texttt{memcached} stand as outliers. Their stealth cache hit rate are 67\% and 85\%, respectively, as they exhibit random page access patterns and high page fault rates. 
In contrast, for Merkle Tree, even with advanced forms of compression and large caches  (1 MB -- 32 KB per core), past work reports low hit rates in the range of 60\%-70\%. Also, while Toleo requires only one stealth version access for a memory write, a Merkle Tree solution traverses the tree until it achieves a cache hit. This requires access at a maximum equal to the number of levels in the tree (10 accesses for a 28 TB memory and 8-ary tree with 3KB root).

\begin{figure*}[ht]
  \centering
  \begin{tabular}{@{}c@{}@{}c@{}}
    \includegraphics[width=0.72\linewidth]{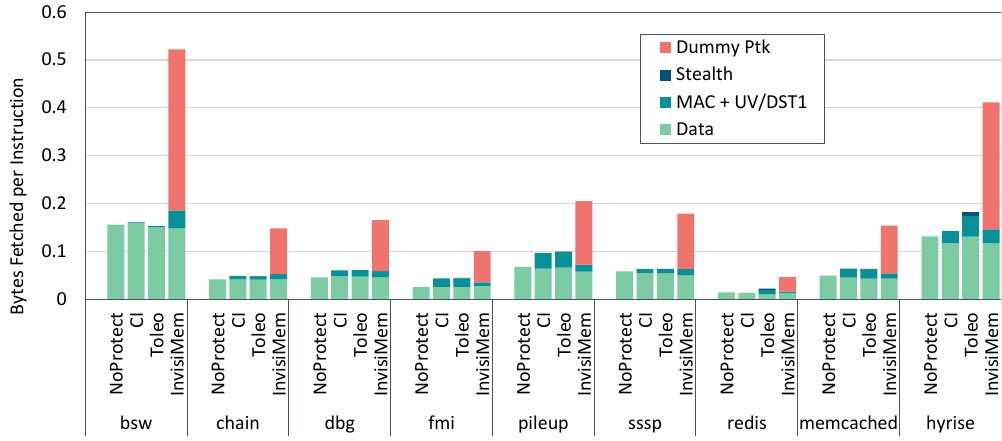} & 
    \includegraphics[width=0.26\linewidth]{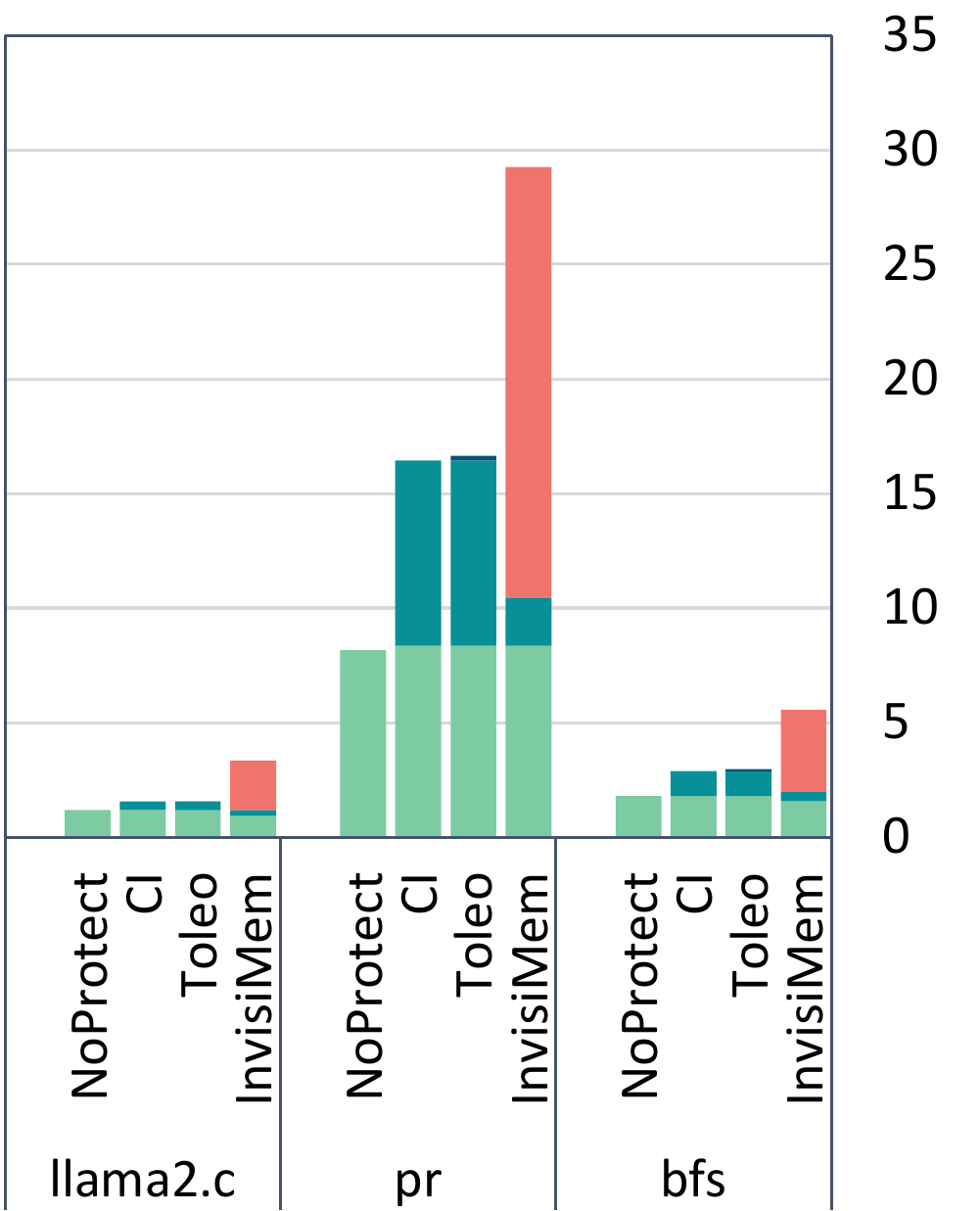} \\
  \end{tabular}
  \caption{Memory bandwidth overhead.}
  \label{fig:access}
\end{figure*}

Performance overhead due to confidentiality (C) is relatively small. Intel TME reports an overhead of less than 2\%\cite{TMEMK, IntelTME} from memory encryption/decryption. CI's performance overhead is primarily due to fetching MACs that are later used for integrity checks. MAC to data ratio for integrity verification is relatively high (1:8). As a result, even with a MAC cache (1 MB) that is 32x larger than our version caches, we observe only an average hit rate of 67\% for MAC accesses. In the worst case, it is as low as 11\%.

The low hit rate for MACs imposes significant pressure on memory bandwidth as shown in Figure~\ref{fig:access}. Each cache block stores eight MACs. Therefore, applications with poor spatial locality ({\tt fmi} and {\tt pr}) suffer from low MAC cache block utilization. Metadata traffic for InvisiMem is lower than CI and Toleo, as the smart memory dynamically groups multiple MACs into one DDR/CXL transaction. Stealth versions, on the other hand, add negligible memory bandwidth overhead compared to data accesses. Even for one of our most memory bandwidth-intensive benchmarks {\tt pr} (page rank), Toleo adds only 2\% additional traffic. This makes it feasible for multiple compute nodes to share the bandwidth of one Toleo smart memory. A low-bandwidth CXL 8x link is sufficient for fetching stealth versions. InvisiMem requires additional memory bandwidth for sending dummy packets to maintain a constant communication rate to mitigate memory address and memory bus timing side-channels.  

\begin{figure}[h]
\centering
\includegraphics[width=\linewidth]{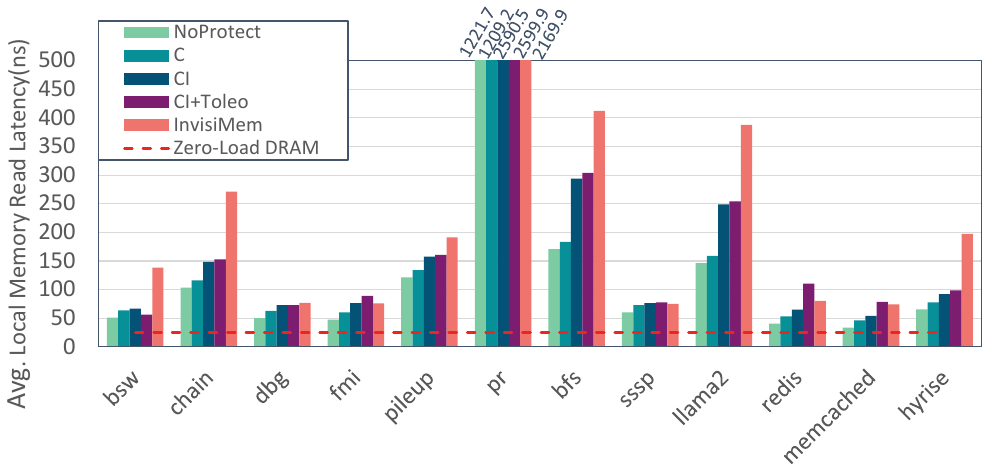}
\caption{Average Memory Read Latency. }
\label{fig:lat}
\end{figure}

Figure \ref{fig:lat} breaks down memory read latency into decryption latency for confidentiality (C), MAC-based integrity check latency (I), and Toleo access latency for freshness (Toleo). AES-XTS decryption adds an 18.6\% overhead. Integrity incurs a higher overhead (36.9\%) due to poor MAC caching. In most cases, Toleo's freshness checks add less than 5\% of read latency, thanks to the high stealth version cache hit rates, except for {\tt redis} and {\tt memcached} which suffers 72\% and 112\% of read latency overhead from freshness checks due to their poor Stealth cache performance. 
Therefore, despite Toleo's remote location in a separate memory drawer, its access latency has minimal impact on the total memory read latency for most workloads. 
InvisiMem suffers an average 2.1x read latency compared to no protection, mainly due to its increased bandwidth pressure, as explained above.

\subsection{Toleo Space Overhead}
\label{sec:eval:meta-date-size}

Table \ref{tab:compression} compares the size of freshness guaranteed versions in the Toleo stealth version representations to the leaf counters in Client SGX\cite{IntelMEE}, VAULT\cite{taassori2018vault} and MorphCtr-128\cite{saileshwar2018morphable}. Toleo achieves an average data-to-stealth version ratio of 240:1, thanks to the common case 341:1 Toleo flat representation and on-demand upgrade to full and uneven. 

\begin{scriptsize}
\begin{table}[h!]
\resizebox{\linewidth}{!}{
\begin{tabular}{|l|l|l|l|}
\hline
\textbf{Representation} & \textbf{\begin{tabular}[c]{@{}l@{}}Trusted Version\\ Rep. Size\end{tabular}}            & \textbf{\begin{tabular}[c]{@{}l@{}}Data Protected \\ per Entry\end{tabular}} & \textbf{\begin{tabular}[c]{@{}l@{}}Data : Version\\ Size Ratio\end{tabular}} \\ \hline
Client SGX (Leaf)\cite{IntelMEE}       & 7B                                                                                      & 64B                                                                          & 9.14:1                                                                  \\ \hline
VAULT (Leaf)\cite{taassori2018vault}            & 64B                                                                                      & 4KB                                                                          & 64:1                                                                    \\ \hline
MorphCtr-128(Leaf)\cite{saileshwar2018morphable}      & 64B                                                                                      & 8KB                                                                          & 128:1                                                                   \\ \hline
Toleo Stealth Flat      & 12B                                                                                      & 4KB                                                                          & 341:1                                                                   \\ \hline
Toleo Stealth Uneven    & 68B\footnotemark{} & 4KB                                                                          & 60:1                                                                    \\ \hline
Toleo Stealth Full      & 228B\footnotemark[\value{footnote}]                                                                                     & 4KB                                                                          & 18:1                                                                    \\ \hline
Toleo Stealth Avg.      & 17.08B                                                                                   & 4KB                                                                          & 240:1                                                                   \\ \hline
\end{tabular}
}
\caption{Freshness Protected Version Size Comparison}
\label{tab:compression}
\end{table}
\end{scriptsize}

\footnotetext{The Toleo flat entry is also used in uneven/full stealth version representations, therefore we account for flat entry size plus uneven/full entry size for these two representations.} 

\begin{figure}[h]
\centering
\includegraphics[width=0.9\linewidth]{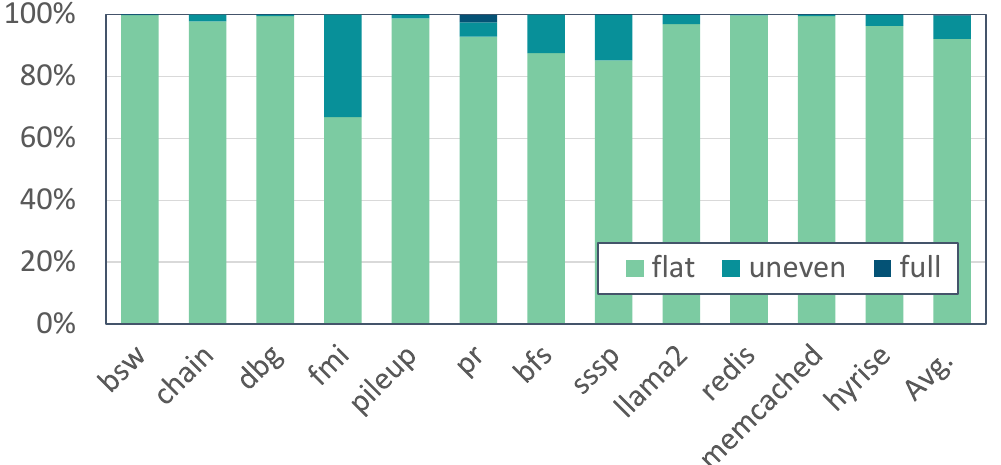}
\caption{Pages classified by their Trip format. }
\label{fig:rep-breakdown}
\end{figure}

We run long simulations with Sniper in cache-only mode to analyze the stealth version representation type breakdown for different data-intensive kernels. As shown in Fig. \ref{fig:rep-breakdown}, a minor 7.5\% of stealth versions are updated to a Toleo uneven format, with a tiny 0.32\% in Toleo full. Dynamic programming ({\tt bsw}, {\tt chain}) and matrix multiplication ({\tt llama2-gen}) kernels enjoy good version locality from their regular memory write patterns, making more than 96\% of their pages flat. Hash table lookups ({\tt dbg}, {\tt pileup}) have irregular read patterns but hash tables are read-only once built. Similarly, key-value store databases ({\tt redis}, {\tt memcached}) are dominated by key lookups. They update their values at relatively low rates. 98\% pages are flat for {\tt dbg}, {\tt pileup}, {\tt redis} and {\tt memcached}. Graph workload ({\tt pr}, {\tt bfs}, {\tt sssp}) are more irregular, making 7-15\% pages uneven or full. The in-memory SQL transaction workload {\tt hyrise} also has irregular writes, but only updates its data structure at write query commits, resulting in 4\% of its pages being uneven. The tree traversal algorithm in {\tt fmi} generates irregular memory updates in the tree structure and lacks version locality, making 33\% of its pages uneven.

\begin{figure}[h]
\centering
\includegraphics[width=0.9\linewidth]{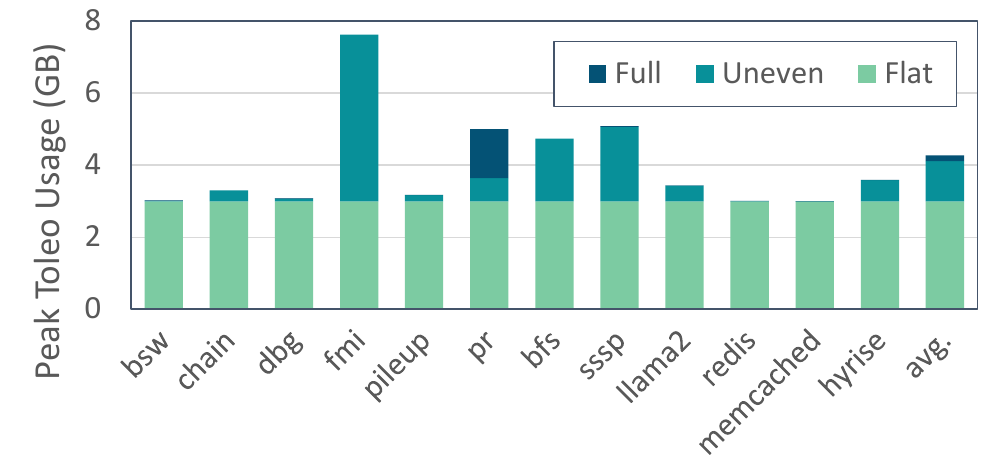}
\caption{Peak Toleo Usage per TB Protected Data}
\label{fig:peak-toleo}
\end{figure}

\begin{figure*}[h]
  \centering
  \begin{tabular}{@{}c@{}@{}c@{}@{}c@{}@{}c@{}@{}c@{}}

    \includegraphics[width=0.19\textwidth]{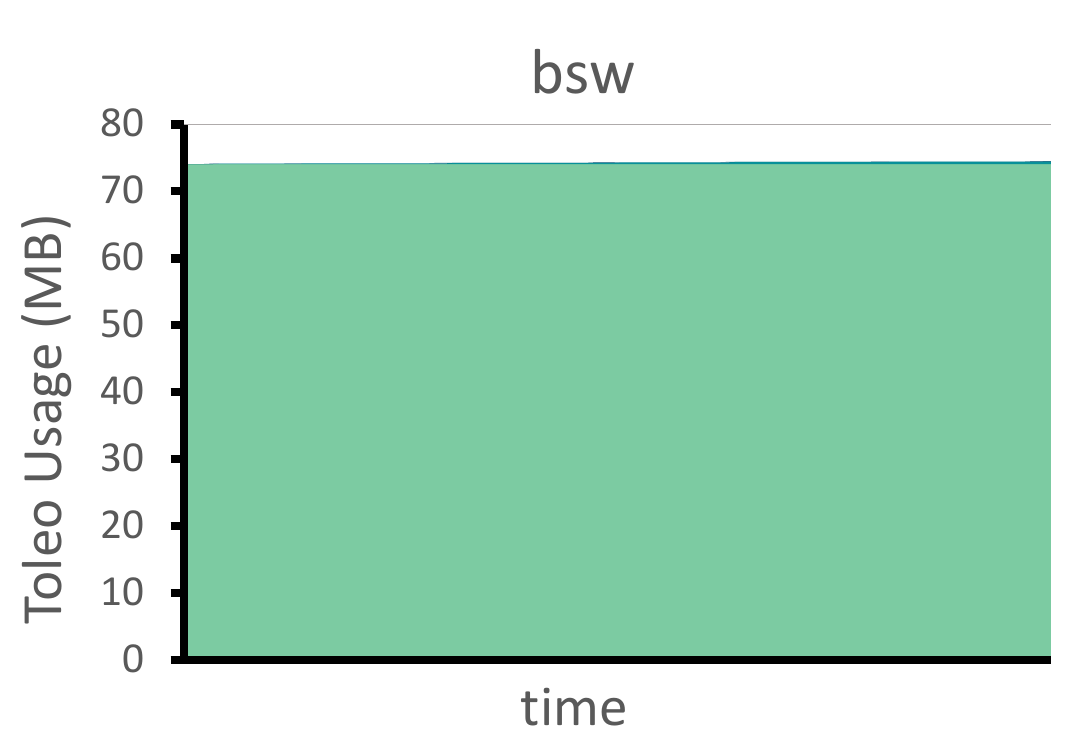} & 
    \includegraphics[width=0.19\textwidth]{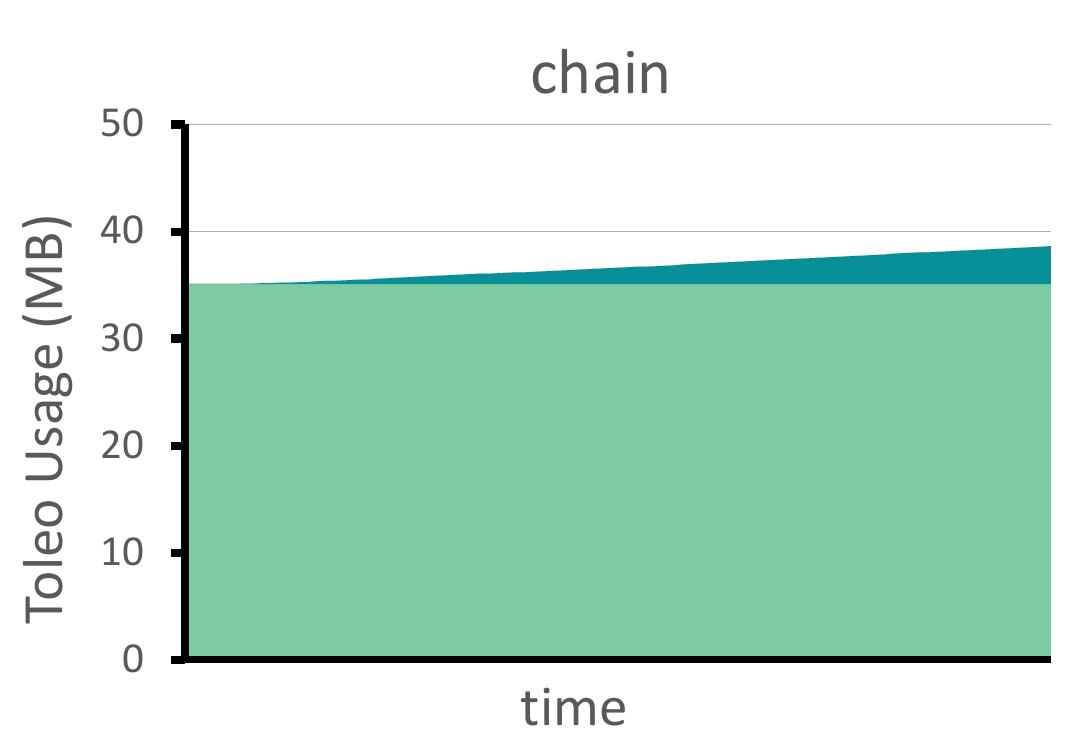} &
    \includegraphics[width=0.19\textwidth]{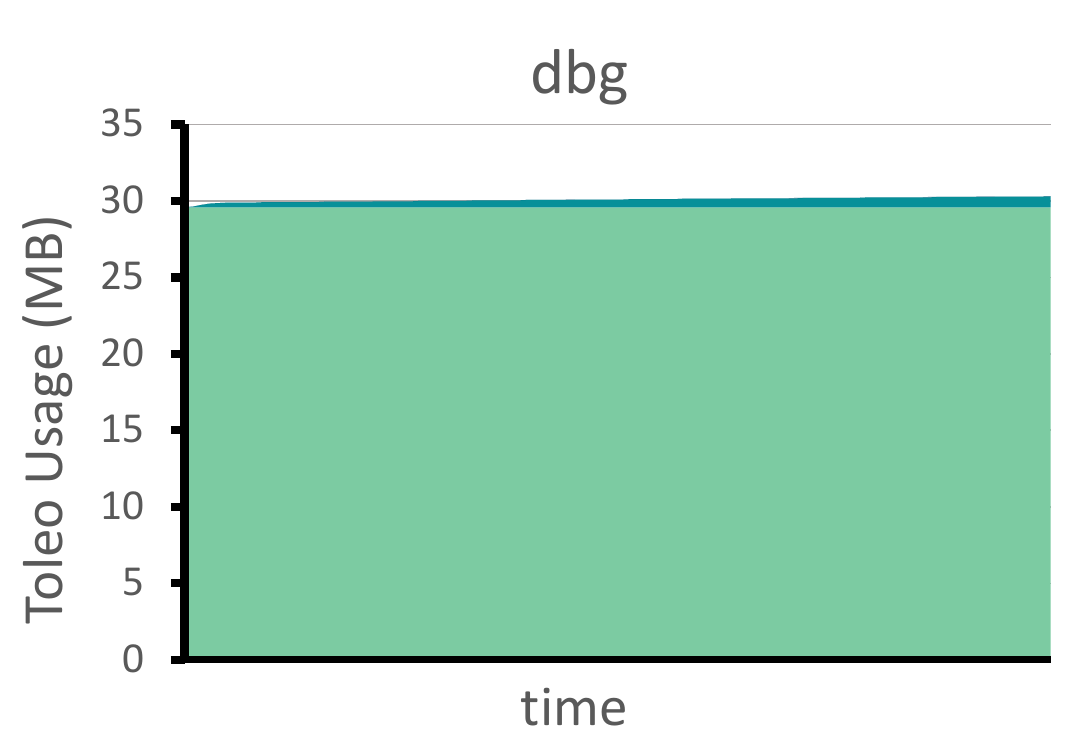} & 
    \includegraphics[width=0.19\textwidth]{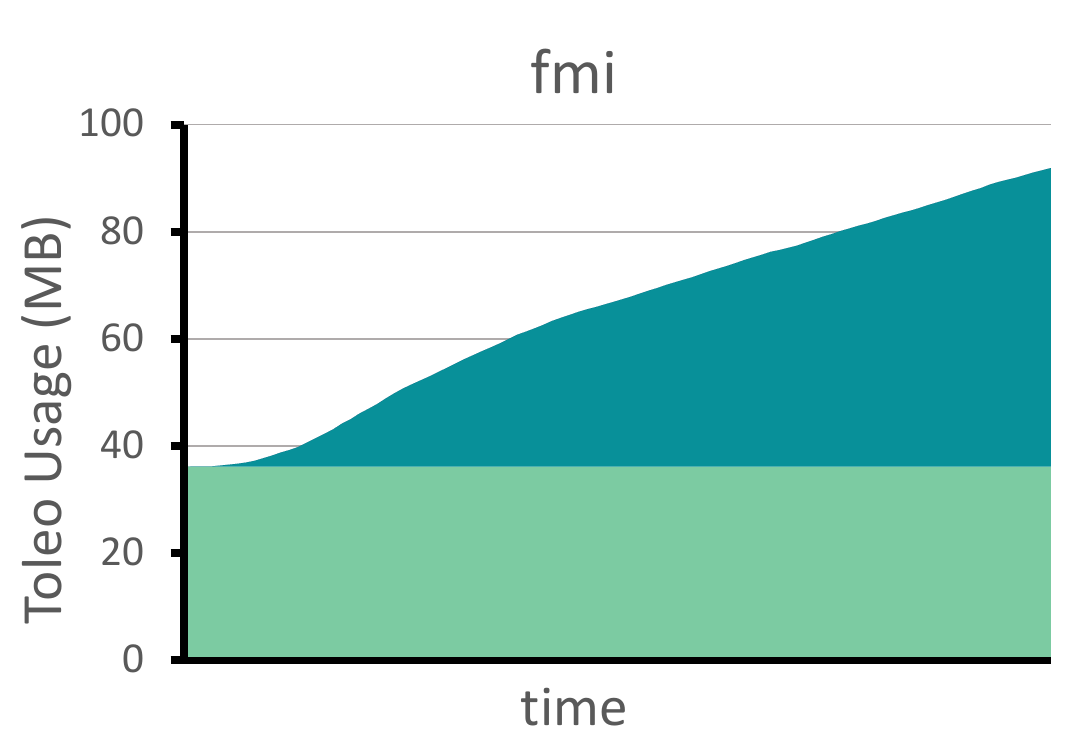} \\
    \includegraphics[width=0.19\textwidth]{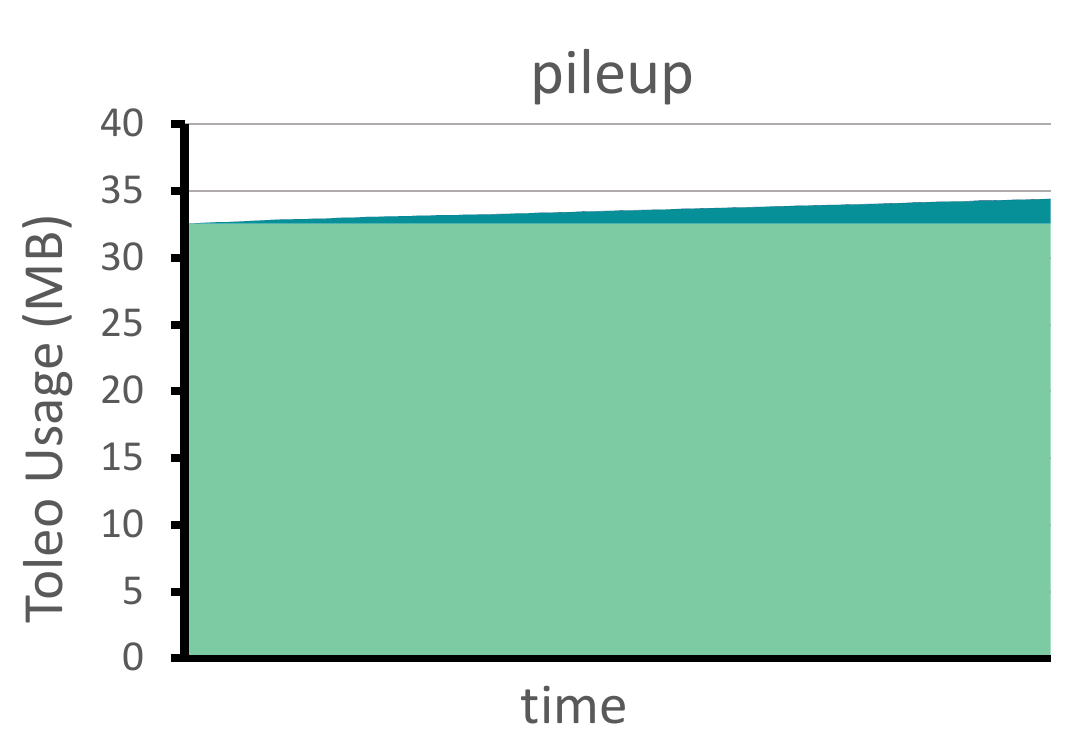} &
    \includegraphics[width=0.19\textwidth]{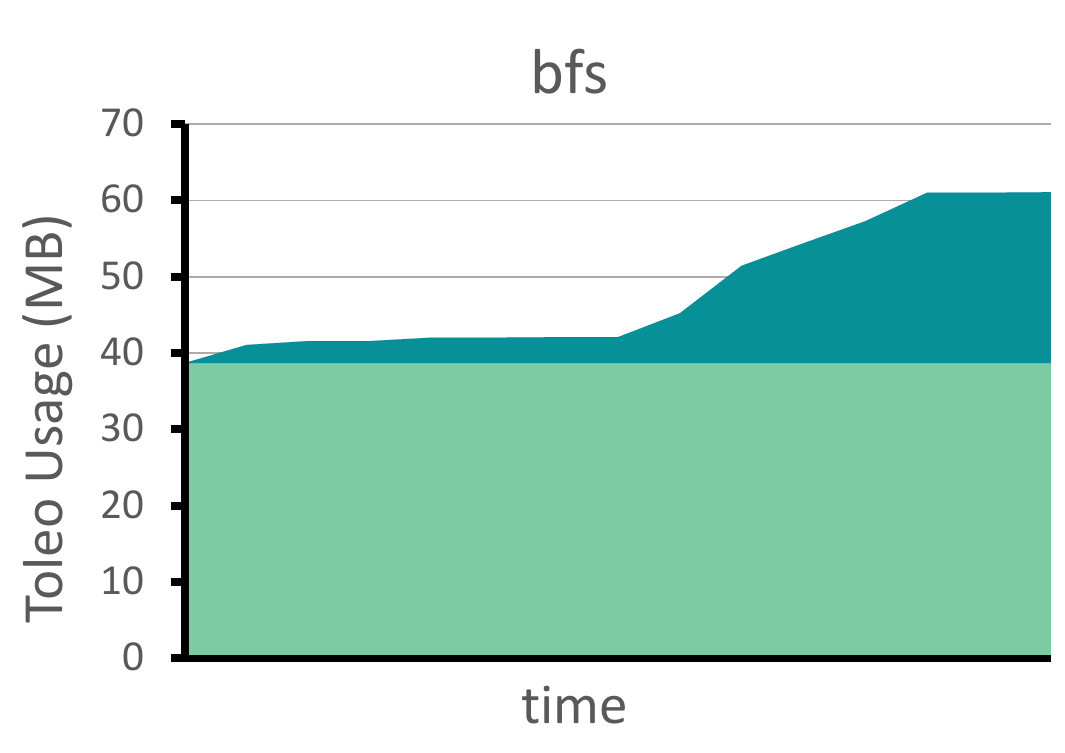} & 
    \includegraphics[width=0.19\textwidth]{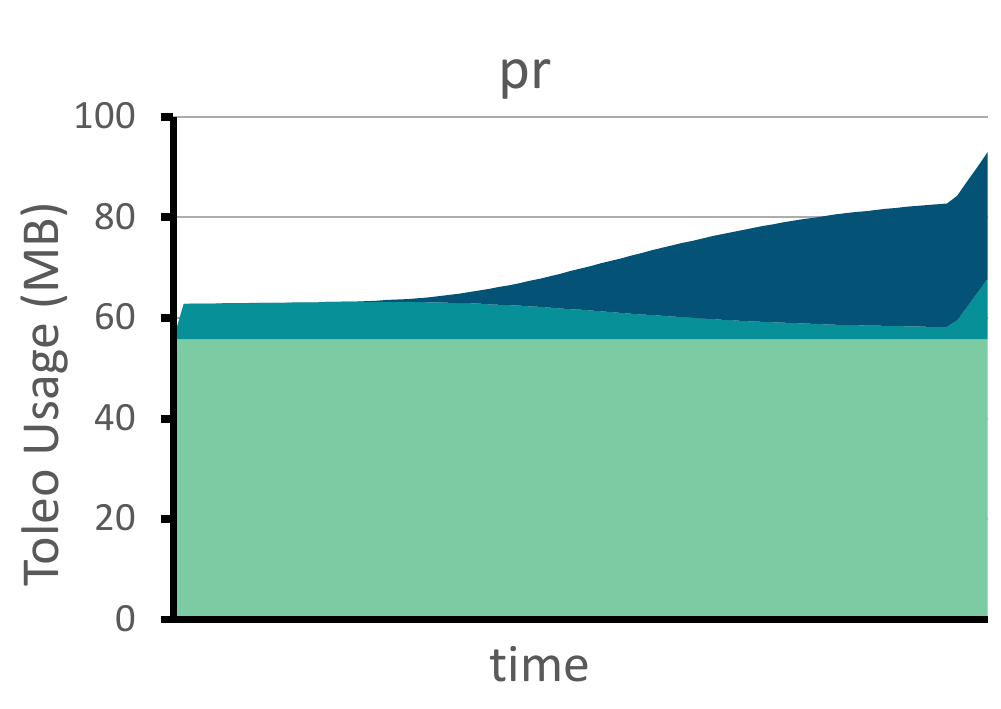} &
    \includegraphics[width=0.19\textwidth]{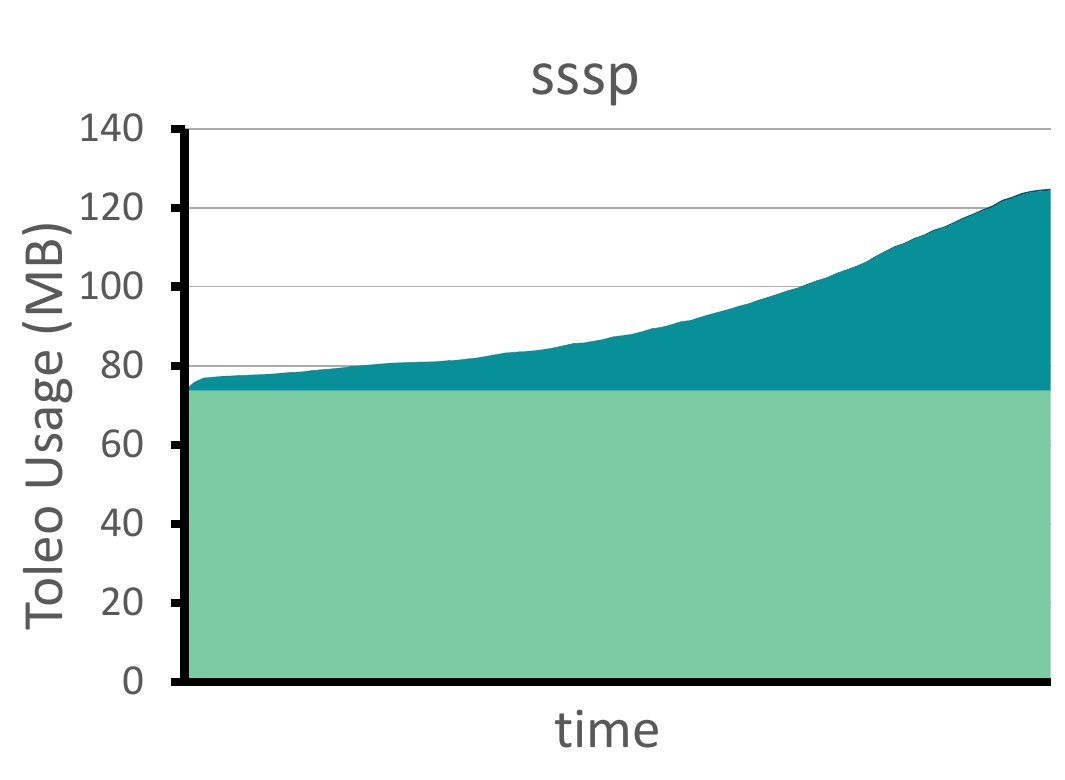} &
    \includegraphics[width=0.1\textwidth]{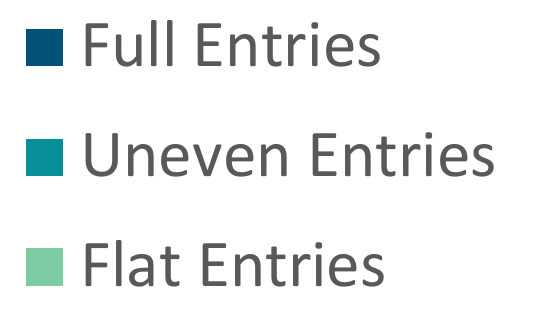}\\
    \includegraphics[width=0.19\textwidth]{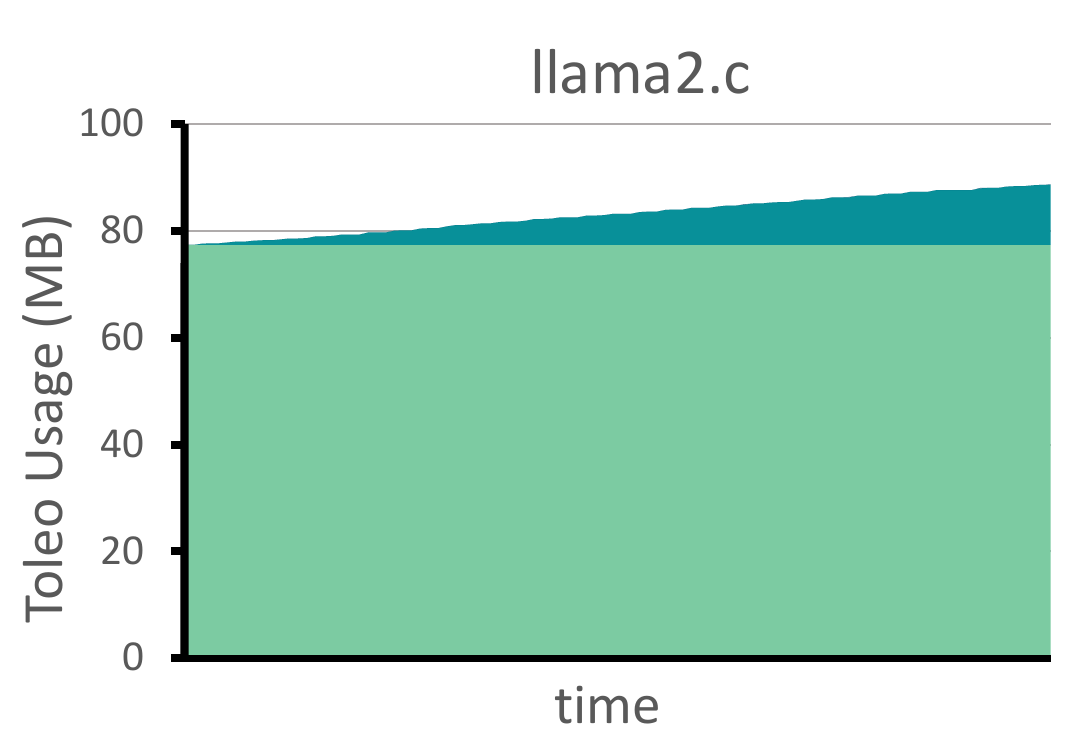}&
    \includegraphics[width=0.19\textwidth]{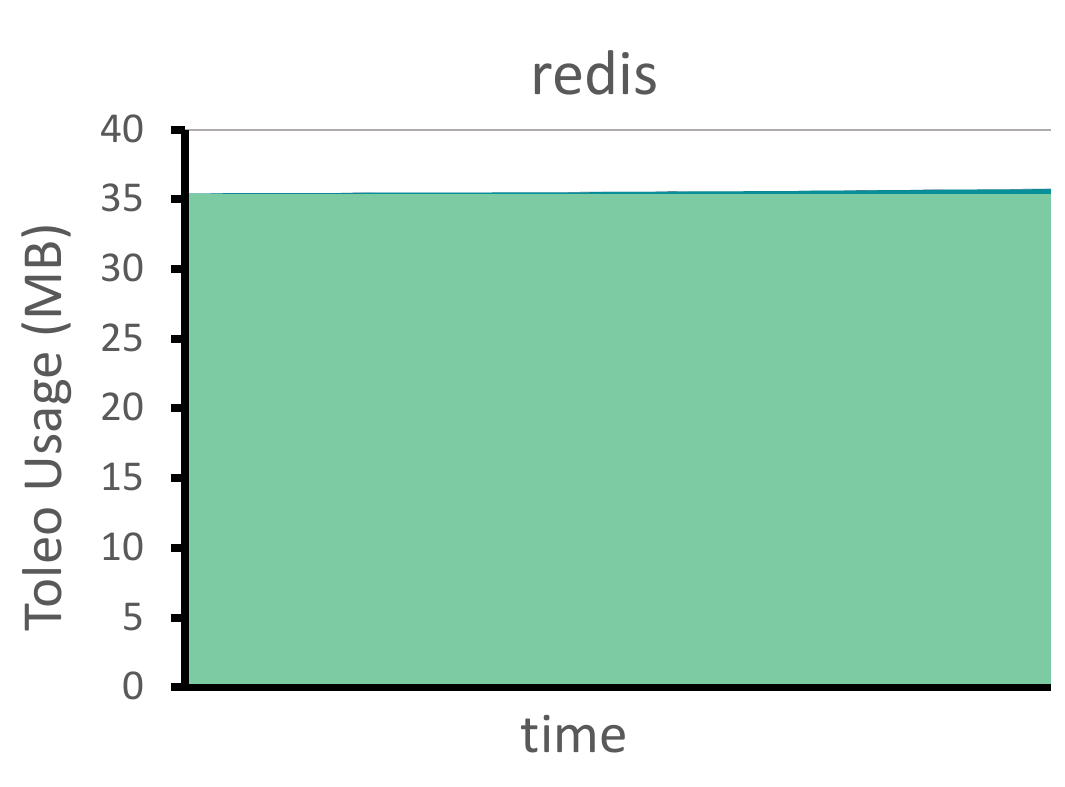}&
    \includegraphics[width=0.19\textwidth]{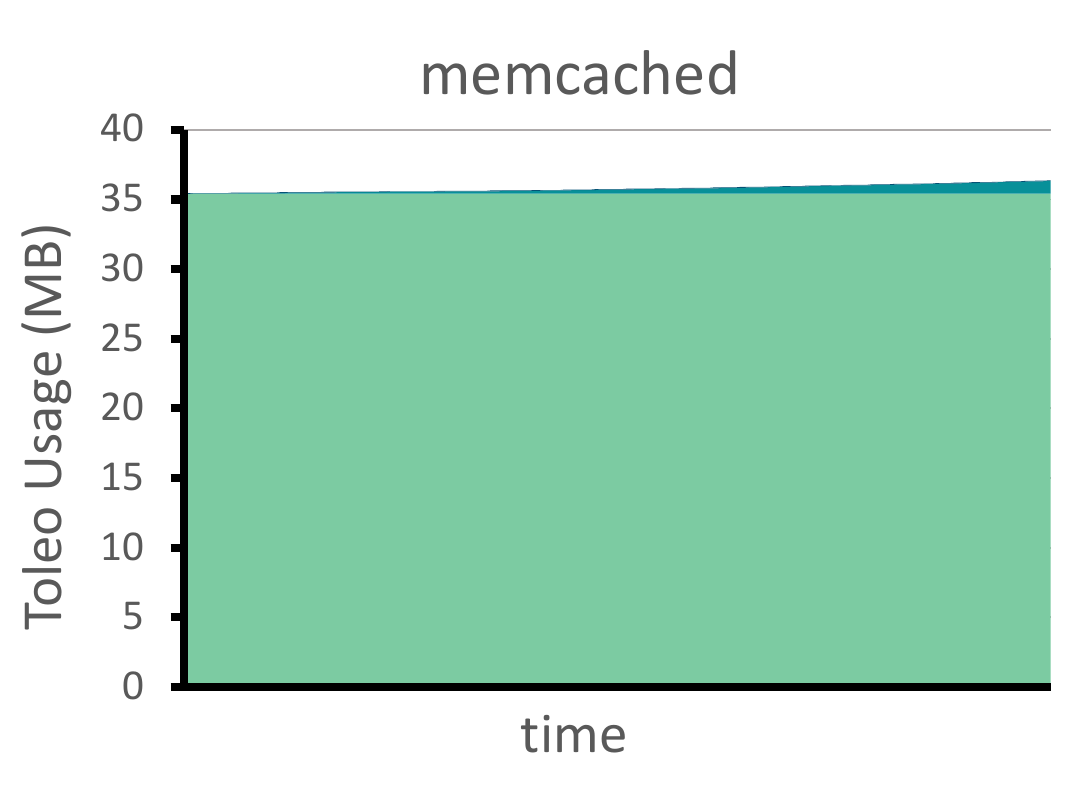}&
    \includegraphics[width=0.19\textwidth]{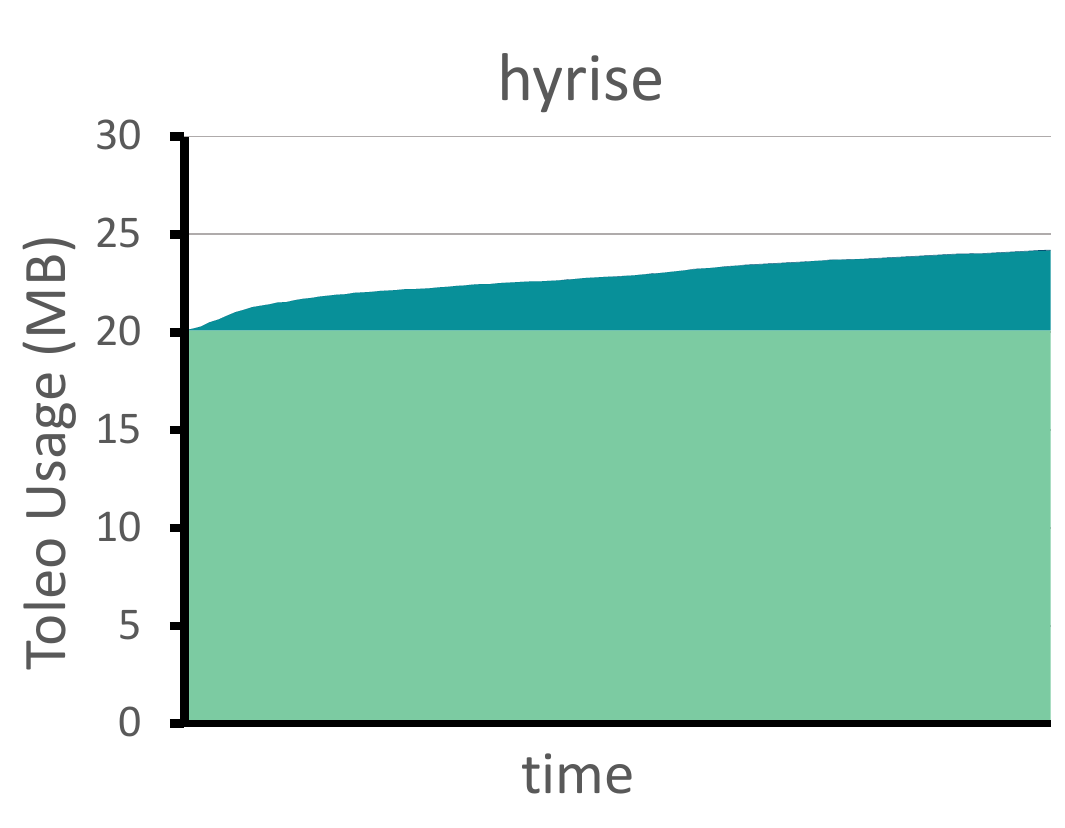}&
    \\
  \end{tabular}

  \caption{Toleo Usage by Trip format w.r.t Time}

  \label{fig:toleousage-bench}
\end{figure*}

Fig. \ref{fig:toleousage-bench} presents the usage of Toleo for various kernels over time. This usage is calculated from static flat entry size derived from kernel RSS plus dynamically allocated uneven/flat entry size reported by the long simulation. Fig. \ref{fig:peak-toleo} reports the peak Toleo usage normalized to RSS, averaging at 4.27 GB per TB-protected data. Most benchmarks (11 out of 12) peak at less than 5.1GB per TB RSS, implying that a 168GB Toleo can protect a 37 TB memory pool without forced entry downgrades. The kernel with the poorest version locality ({\tt fmi}) needs 7.6GB of Toleo per TB RSS. Such a kernel should be collocated with good version locality workloads on the same Toleo, such as {\tt bsw}, {\tt chain}, {\tt dbg}, {\tt pileup} or {\tt llama2-gen}.

\subsection{Area and Power Overhead}

The Toleo system reduces the area overhead for caching Merkle Tree and full version counters on the host processor. Instead, the L2 TLB stealth version extension increases the TLB data size from 1.5KB to 4.5K but does not change its replacement policy and tag array. The stealth version overflow buffer of 28KB is shared by 32 cores, which is only 2.7\% the size of the MAC cache. Moreover, only 1\% of the bytes fetched off-chip is for freshness verification, leading to considerable data movement energy savings.

\section{Related Work}
\label{sec:related}
This paper is the first to propose the use of smart memory for storing version numbers and leveraging CXL to provide freshness at scale. Prior works have used smart memory to defend against various side-channels in computing hardware: thermal~\cite{gu3dsecurity} and cache~\cite{valamehr3Dcontrolplane}, and address~\cite{aga2017invisimem, awad2017obfusmem}.  Invisimem~\cite{aga2017invisimem} and Obfusmem~\cite{awad2017obfusmem} solve address side-channel in trusted hardware by encrypting memory address during transit from the processor to the memory and decrypting it using logic inside of the secure smart memory. In contrast, Toleo's goal is freshness. Unlike Toleo, they store the entire data in smart memory, which is more expensive than conventional memory. We further reduce cost by sharing Toleo with multiple nodes in a rack server. 

Several prior works have exploited version locality~\cite{umar2022softvn,leetnpu,nacommoncounters,hua2022mgx,taassori2018vault,saileshwar2018morphable}. Recent works seek to compress version numbers in the Merkle tree based on version locality. VAULT~\cite{taassori2018vault} has a variable arity for each node in the Merkle tree depending on the version locality and can fit 16-64 version numbers in a cache line. Morphable Counters~\cite{saileshwar2018morphable} improves on this by dynamically changing how version numbers are compressed depending on the workload. For example, in the case where one version number is written far more than the version numbers surrounding it, there is a representation that tracks small version numbers with a bit vector while reserving the rest of the bits for the larger version number. Both of these works improve the arity of the Merkle tree and improve performance. However, they are limited to storing version numbers at the cache line granularity. Toleo eliminates the need for a Merkle tree which creates an opportunity for more flexible and space-efficient version number compression representations. These optimizations also enable efficient caching versions by extending the last-level TLB. 

SoftVN~\cite{umar2022softvn} has developers annotate data structures that have uniform write patterns to assign version numbers at the data structure granularity. Similarly, TNPU~\cite{leetnpu} tracks version numbers at the software-defined granularity, assigning a version number to each tensor object for NPU applications. Toleo does not pass the burden of managing version numbers onto developers, as our hardware design automatically exploits version locality.

Na et al. \cite{nacommoncounters} scan GPU memory for uniformly written regions of memory and use a "common counter" that is cached on the processor. Lastly, MGX~\cite{hua2022mgx} takes advantage of the statically determined data access patterns in accelerator applications to predetermine what the version number should be during execution, eliminating the need to store version numbers in memory completely. Toleo tracks locality at the granularity of a page. This is significant because prior works depend on accelerator-specific access patterns and granularities, which we do not. Additionally, tracking at the page granularity allows us to take advantage of shared last-level TLB caching~\cite{bhattacharjee2011tlb} without adding additional caches for version numbers or polluting data caches~\cite{nacommoncounters,gassend2003caches,mapscaching,lee2016reducing}.

Lastly, several works improve performance by optimizing Merkle tree traversal. This can be achieved by parallelizing operations in the tree~\cite{hall2005pat,elbaz2007tec}, skewing the tree depth to favor frequently accessed paths~\cite{skewedTree}, and caching schemes to improve Merkle tree hit rate~\cite{gassend2003caches,mapscaching,lee2016reducing}. We obviate the need for these optimizations due to not using a Merkle tree to protect freshness. Another way to improve freshness checks is to assume it will always succeed and speculatively execute until the check is completed~\cite{lehman2016poisonivy,shi2006spec}. This is orthogonal to our contribution and could be used in conjunction with Toleo to further reduce latency in our system, however, there are many works demonstrating vulnerabilities associated with speculative execution in trusted hardware~\cite{kocher2020spectre,sgxpectre,van2018foreshadow} that leads us to believe this optimization may be dangerous.

\section{Conclusion}
Freshness is important for realizing zero-trust solutions. Unfortunately, as Merkle tree-based solutions do not scale to terabyte-scale memory, even with compression and caching optimizations, industry is forgoing this security feature.  We introduce a low-cost method using smart memory and CXL technologies, coupled with version space optimizations, to secure freshness for 28 TB of main memory. This solution incurs only a 1\% performance overhead and a modest on-chip area increase (31 KB total for a 32-core processor).
\label{revis:smartNIC}
The TCB of confidential computing is expanding beyond the traditional processor boundaries. Recently, Nvidia released the Grace Hopper GPU which includes features that enable confidential computing in an accelerator~\cite{nvidia_gracehopper}.  Our work is another step in this direction as it explores modern smart memory devices and how their integrated logic can play a role in confidential computing. Smart NICs~\cite{zhou2024smartnic,nvidia_bluefield} (e.g., NVIDIA Bluefield) include isolated memory and accelerators for cryptographic operations. Future work could investigate their role in enabling confidential computing. 
\begin{acks}
We thank Yu Hua and anonymous reviewers for their comments which helped improve this paper. This project was supported in part by the Kahn Foundation and NSF award \#2403119. 
\end{acks}
\appendix

\section{Artifact Appendix}

\subsection{Abstract}
This Artifact Appendix describes the generation of Toleo performance metrics, including runtime, memory latency (see Sec. \ref{sec:performance}), and Toleo space usage (see Sec. \ref{sec:eval:meta-date-size}). The reported results are from execution-driven cycle-accurate simulations of benchmarks described in sec. \ref{sec:evaluation}. A successful simulation will generate performance statistics including runtime, CPI, cache hit rate, memory latency, and memory bandwidth utilization. Additionally, the simulation also generates Trip format stats describing how many pages are stored in which format. 

We provide a GitHub repository containing the Toleo simulator source code, a list of forked benchmarks used in Section 7.2, an automated workflow script, and a Docker environment. A x86-64 Linux host with at least 32 cores, 64GB of RAM, and 512GB of free disk space is required to run the simulation.

\subsection{Artifact check-list (meta-information)}

{\small
\begin{itemize}
\item {\bf Simulator:} sniper-toleo.
\item {\bf Benchmarks:} {\tt genomicsbench}, {\tt gapbs}, {\tt llama2.c}, {\tt redis},\\ {\tt memcached}, {\tt hyrise}, {\tt memtier\_benchmark}.
\item {\bf Run-time environment:} Depends on i386 architecture packages {\tt libc6
}, {\tt libncurses5
}, and {\tt libstdc++6
}.
\item {\bf Hardware:} CPU with 32+ cores, 64GB of memory, and 128GB of free disk space.
\item {\bf Simulation Time:} 1.5 hours to 3 days per test case per setup. Toy benchmarks (<20 minutes simulation time) are provided.
\item {\bf Output:} Performance stats in {\tt sim.out} and Trip stats in {\tt dram\_trace-analysis.csv}.
\item {\bf Time needed to prepare workflow (approx.):} 5 hours.
\item {\bf Publicly available?:} Yes.
\item {\bf Code licenses (if publicly available):} MIT License.
\item {\bf Archived (provide DOI):} No.
\end{itemize}
}

\subsection{Description}

\subsubsection{How to access}

Clone from github repo: \url{https://github.com/joydddd/sniper-toleo}

\subsubsection{Hardware dependencies}
Machine with more than 32 cores and more than 64GB of memory.

\subsection{Installation}

Please refer to {\tt README.md} from this repo: \url{https://github.com/joydddd/sniper-toleo} for detailed instructions. We recommend installing and running the simulation in our provided Docker container.
Briefly, users should first build the DRAM simulator DRAMsim3, and then build the sniper-toleo simulator. A successful installation of the sniper-toleo simulator enables the simulation of multi-threaded benchmarks running on a Toleo-protected memory system. A test program is provided in \verb|test/fft|. 

\subsection{Experiment customization}
Sniper-toleo can simulate any CPU benchmarks of fewer than 32 threads on our evaluated setups: memory protected by Toleo, C+I, no protection and invisimem as descripted in Sec. \ref{sec:evaluation}. Please refer to the "Run Your Own Benchmark"  section in {\tt README.md} for detailed instructions. 

\subsection{Experiment Workflow}
Users can follow the instructions in {\tt BenchSetup.md} to clone and build CPU benchmarks found in this list of forks: \url{https://github.com/stars/joydddd/lists/toleo}. 

We provide a test automation script {\tt run\_toleo\_sim.py} to batch run simulations on the above benchmarks. It also allows users to debug the workflow by running the benchmarks natively or with PIN instrumentation but without simulation. Please run {\tt ./run\_toleo\_sim.py --help } or refer to {\tt README.md} for details.

\subsection{Evaluation and expected results}
After a successful simulation, users should find performance metrics in {\tt sim.out} and Trip format stats in {\tt dram\_trace\_} {\tt analysis.csv}. Default output path and example outputs are provided in {\tt README.md}.

\bibliographystyle{plain}
\bibliography{main}

\end{document}